\documentstyle[12pt,psfig]{article}
\makeatletter
\def\@cite#1#2{{[{#1}]\if@tempswa\typeout
{IJCGA warning: optional citation argument
ignored: `#2'} \fi}}

\newcount\@tempcntc
\def\@citex[#1]#2{\if@filesw\immediate\write\@auxout{\string\citation{#2}}\fi
  \@tempcnta\z@\@tempcntb\m@ne\def\@citea{}\@cite{\@for\@citeb:=#2\do
    {\@ifundefined
       {b@\@citeb}{\@citeo\@tempcntb\m@ne\@citea\def\@citea{,}{\bf ?}\@warning
       {Citation `\@citeb' on page \thepage \space undefined}}%
    {\setbox\z@\hbox{\global\@tempcntc0\csname b@\@citeb\endcsname\relax}%
     \ifnum\@tempcntc=\z@ \@citeo\@tempcntb\m@ne
       \@citea\def\@citea{,}\hbox{\csname b@\@citeb\endcsname}%
     \else
      \advance\@tempcntb\@ne
      \ifnum\@tempcntb=\@tempcntc
      \else\advance\@tempcntb\m@ne\@citeo
      \@tempcnta\@tempcntc\@tempcntb\@tempcntc\fi\fi}}\@citeo}{#1}}
\def\@citeo{\ifnum\@tempcnta>\@tempcntb\else\@citea\def\@citea{,}%
  \ifnum\@tempcnta=\@tempcntb\the\@tempcnta\else
   {\advance\@tempcnta\@ne\ifnum\@tempcnta=\@tempcntb \else \def\@citea{--}\fi
    \advance\@tempcnta\m@ne\the\@tempcnta\@citea\the\@tempcntb}\fi\fi}
\makeatother
\newenvironment{Eqnarray}%
     {\arraycolsep 0.14em\begin{eqnarray}}{\end{eqnarray}}

\def\simlt{\stackrel{<}{{}_\sim}}
\def\simgt{\stackrel{>}{{}_\sim}}
\def\be{\begin{equation}}
\def\ee{\end{equation}}
\def\bear{\be\begin{array}}
\def\eear{\end{array}\ee}
\def\bea{\begin{Eqnarray}}
\def\eea{\end{Eqnarray}}

\def\lsim{\mathrel{\raise.3ex\hbox{$<$\kern-.75em\lower1ex\hbox{$\sim$}}}}
\def\gsim{\mathrel{\raise.3ex\hbox{$>$\kern-.75em\lower1ex\hbox{$\sim$}}}}
\def\ifmath#1{\relax\ifmmode #1\else $#1$\fi}
\def\ls#1{\ifmath{_{\lower1.5pt\hbox{$\scriptstyle #1$}}}}

\def\beq{\begin{equation}}
\def\eeq{\end{equation}}
\def\beqa{\begin{Eqnarray}}
\def\eeqa{\end{Eqnarray}}

\def\baselinestretch{1}
\begin{document}
\def\IJMPA #1 #2 #3 {{\sl Int.~J.~Mod.~Phys.}~{\bf A#1}\ (19#2) #3$\,$}
\def\MPLA #1 #2 #3 {{\sl Mod.~Phys.~Lett.}~{\bf A#1}\ (19#2) #3$\,$}
\def\NPB #1 #2 #3 {{\sl Nucl.~Phys.}~{\bf B#1}\ (19#2) #3$\,$}
\def\PLB #1 #2 #3 {{\sl Phys.~Lett.}~{\bf B#1}\ (19#2) #3$\,$}
\def\PR #1 #2 #3 {{\sl Phys.~Rep.}~{\bf#1}\ (19#2) #3$\,$}
\def\JHEP #1 #2 #3 {{\sl JHEP}~{\bf #1}~(19#2)~#3$\,$}
\def\PRD #1 #2 #3 {{\sl Phys.~Rev.}~{\bf D#1}\ (19#2) #3$\,$}
\def\PTP #1 #2 #3 {{\sl Prog.~Theor.~Phys.}~{\bf #1}\ (19#2) #3$\,$}
\def\PRL #1 #2 #3 {{\sl Phys.~Rev.~Lett.}~{\bf#1}\ (19#2) #3$\,$}
\def\RMP #1 #2 #3 {{\sl Rev.~Mod.~Phys.}~{\bf#1}\ (19#2) #3$\,$}
\def\ZPC #1 #2 #3 {{\sl Z.~Phys.}~{\bf C#1}\ (19#2) #3$\,$}
\def\PPNP#1 #2 #3 {{\sl Prog. Part. Nucl. Phys. }{\bf #1} (#2) #3$\,$}

\catcode`@=11
\newtoks\@stequation
\def\subequations{\refstepcounter{equation}%
\edef\@savedequation{\the\c@equation}%
  \@stequation=\expandafter{\theequation}
  \edef\@savedtheequation{\the\@stequation}
  \edef\oldtheequation{\theequation}%
  \setcounter{equation}{0}%
  \def\theequation{\oldtheequation\alph{equation}}}
\def\endsubequations{\setcounter{equation}{\@savedequation}%
  \@stequation=\expandafter{\@savedtheequation}%
  \edef\theequation{\the\@stequation}\global\@ignoretrue

\noindent}
\catcode`@=12
\begin{titlepage}

\title{{\bf Nearly degenerate neutrinos, Supersymmetry and radiative
corrections}}
\vskip2in
\author{ 
{\bf J.A. Casas$^{1,2}$\footnote{\baselineskip=16pt E-mail: {\tt
casas@mail.cern.ch}}}, 
{\bf J.R. Espinosa$^{2}$\footnote{\baselineskip=16pt E-mail: {\tt
espinosa@mail.cern.ch}. }
\footnote{\baselineskip=16pt
On leave of absence from Instituto de Matem\'atica
y F\'{\i}sica Fundamental, CSIC, Madrid (Spain)}}, 
{\bf A. Ibarra$^{1}$\footnote{\baselineskip=16pt  E-mail: {\tt
alejandro@makoki.iem.csic.es}}} and 
{\bf I. Navarro$^{1}$\footnote{\baselineskip=16pt Email: {\tt
ignacio@makoki.iem.csic.es}}}\\ 
\hspace{3cm}\\
 $^{1}$~{\small Instituto de Estructura de la materia, CSIC}\\
 {\small Serrano 123, 28006 Madrid}
\hspace{0.3cm}\\
 $^{2}$~{\small Theory Division, CERN}\\
{\small CH-1211 Geneva 23, Switzerland}.
}
\date{}
\maketitle
\def\baselinestretch{1.15}
\begin{abstract}
\noindent
If neutrinos are to play a relevant cosmological r\^ole, they must be
essentially degenerate with a mass matrix of the bimaximal mixing type. 
We study this scenario in the MSSM framework, finding that if neutrino
masses are produced by a see-saw mechanism, the radiative corrections give
rise to mass splittings and mixing angles that can accommodate the
atmospheric and the (large angle MSW) solar neutrino oscillations.  This
provides a natural origin for the $\Delta m^2_{sol} \ll \Delta m^2_{atm}$
hierarchy.  
On the other hand, the vacuum oscillation solution to the
solar neutrino problem is always excluded. 
We discuss also in the SUSY scenario other possible effects
of radiative corrections involving the new neutrino Yukawa couplings,
including implications for triviality limits on the Majorana mass, the
infrared fixed point value of the top Yukawa coupling, and gauge coupling
and bottom-tau unification.
 \end{abstract}

\thispagestyle{empty}
\leftline{}
\leftline{CERN-TH/99-142}
\leftline{May 1999}
\leftline{}

\vskip-21cm
\rightline{}
\rightline{IEM-FT-193/99}
\rightline{CERN-TH/99-142}
\rightline{hep-ph/9905381}
\vskip3in

\end{titlepage}
\setcounter{footnote}{1} \setcounter{page}{1}
\newpage
\baselineskip=20pt

\noindent

\section{Introduction}

If neutrinos are to play a relevant cosmological role, their masses
should be ${\cal O}$(eV). In that case, since atmospheric and solar neutrino
anomalies~\cite{SK} indicate that mass-squared splittings are at most
$10^{-2}\ {\rm eV}^2$, neutrinos must be almost
degenerate~\cite{GG}-\cite{degenerofilia}.  On the other hand,
supersymmetry (SUSY) is a key ingredient in most of the extensions of
the Standard Model (SM) which are candidate for a more fundamental
theory.  In this paper we will analyze, within the supersymmetric
framework,  under which circumstances the ``observed'' mass splittings
between quasi-degenerate neutrinos arise naturally (or not), as a
radiative effect, in
agreement  with all the available experimental data.

This problem has been also addressed in a recent paper by Ellis and
Lola \cite{EL}, in which they treat the neutrino mass matrix, ${\cal
M}_\nu$, as an effective operator, emerging at some scale, $\Lambda$,
with the bimaximal mixing form. Then the renormalization group (RG)
analysis shows that the splittings and mixings at low energy are not
in agreement with observations. Here we take a more general point
of view. Besides exploring the effective operator scenario, we focus
our attention in the (well motivated) case in which this operator is
produced by a see-saw mechanism\footnote{An alternative possibility to
generate small non-zero neutrino masses in the MSSM involves
$R$-parity breaking \cite{Rp}. See Ref.~\cite{goran} for a discussion on the
interplay between the two possibilities.}. This introduces crucial
differences in the analysis. In particular, the form of ${\cal M}_\nu$
is modified by a first stage of  RG running of
the neutrino Dirac-Yukawa matrix and the  right-handed neutrino mass
matrix from the high energy scale (say $M_p$ or $M_{GUT}$) to $\Lambda$,
which is identified with the Majorana mass scale. As we will see, this
modification allows in many cases to
reconcile the scenario with experiment, providing also a natural
origin for the ``observed'' solar-atmospheric hierarchy of splittings,
$\Delta m^2_{sol}\ll\Delta m^2_{at}$.

In a recent paper~\cite{cein}, we performed a similar analysis in the SM
framework (in which the only particles added to the SM are three
right-handed neutrinos), also with positive results. There are similarities
and differences between the SUSY and the SM cases. 
First, SUSY introduces additional unknowns in the scenario, particularly
the supersymmetric mass spectrum. In the analysis, the only r\^ole of this
spectrum is to give the threshold scale(s)
below which the effective theory is just the SM. In this sense, the
combination of (negative) experimental data and naturalness requires
$M_{SUSY}\sim 1$ TeV. Of course, there may be appreciable differences
between the masses of various supersymmetric particles (squarks,
gluino, charginos, etc.). Still, the variation in the results of the RG
analysis are not important. Hence, we will take a unique threshold at
$M_{SUSY}= 1$ TeV throughout the paper. Let us mention here that, apart
from three right-handed neutrinos, we will assume a minimal spectrum of
particles; in other
words we will work within the minimal supersymmetric standard model
(MSSM).
Second (and more important), in the
supersymmetric regime, the charged-lepton
Yukawa couplings are multiplied by a factor $1/\cos\beta$ with respect to
their SM value.
These couplings (together with the neutrino Yukawa
couplings)
play a major r\^ole in the radiative  modification of the form
of ${\cal M}_\nu$. Thus, the results are going to present a strong
dependence on
$\tan\beta$ (we recall that $\tan\beta$ is defined as the ratio of the
expectation values of the two supersymmetric Higgs doublets,
$\tan\beta\equiv \langle H_2^0\rangle/\langle H_1^0\rangle$).
Finally, the renormalization group equations (RGEs) themselves are
different in the SUSY and in the SM cases. The difference is not just
quantitative (i.e. differences in the size of the various
coefficients), but also qualitative. In particular, the modification
of the ${\cal M}_\nu$ texture due to the contribution from the
charged-lepton Yukawa couplings has opposite signs in the two cases
(while the contribution from the neutrino Yukawa couplings themselves 
remains with the same sign).

\vspace{0.2cm} Let us briefly review the current relevant experimental
constraints on neutrino masses and mixing angles (a more detailed
account is given in ref.~\cite{cein}). Observations of atmospheric
neutrinos are well described by  $\nu_\mu-\nu_\tau$ oscillations driven by
a mass
splitting and a mixing angle in the range \cite{range}
\bea 
5\times10^{-4}\ {\rm eV}^2 <  &\Delta m^2_{at}& < 10^{-2}\ {\rm
eV}^2\ , \nonumber\\ \sin^22\theta_{at}&>&0.82\ .
\label{atm}
\eea
Concerning the solar neutrino problem, as has been shown in
ref.~\cite{GG} the small angle MSW solution is unplausible in a
scenario of nearly degenerate neutrinos, so we are left with the large
angle MSW (LAMSW) and the vacuum oscillation (VO) solutions, which
require
mass splittings and mixing angles in the following
ranges

\vspace{0.2cm} 
LAMSW solution:
\bea 10^{-5}\ {\rm eV}^2 <  &\Delta m^2_{sol}& < 2\times 10^{-4}\ {\rm
eV}^2, \nonumber\\ 0.5 < &\sin^22\theta_{sol}& < 1.
\label{LAMSW}
\eea

\vspace{0.2cm} 
VO solution:
\bea 5\times10^{-11}\ {\rm eV}^2 <  &\Delta m^2_{sol}& <
1.1\times10^{-10}\ {\rm eV}^2, \nonumber\\ \sin^22\theta_{sol} &>& 0.67\,
.
\label{VO}
\eea
From the previous equations, it is apparent the hierarchy of  mass
splittings between the different species of neutrinos, $\Delta
m^2_{sol}\ll\Delta m^2_{at}$, which  should be reproduced by any
natural explanation of those splittings.  Let us also remark that it
is not clear at the moment the value of the upper bound on
$\sin^22\theta_{sol}$ [see eq.(\ref{LAMSW})]. As we will see, an upper
limit like $\sin^22\theta_{sol}<0.99$ or even greater, may disallow
the scenario examined in this paper.  For the moment, we will not consider
any upper
bound on $\sin^22\theta_{sol}$ (see ref.\cite{cein} for a more
detailed discussion). On the other hand,  according to the most recent 
combined analysis of SK $+$ CHOOZ data (last paper of ref.~\cite{range})
the third independent angle, say $\phi$ (the one mixing the electron 
with the most split mass eigenstate), is constrained to have low values,
$\sin^22\phi<0.36\, (0.64)$ at 90\% (99\%) C.L.

Other relevant experimental information concerns the non-observation of
neutrinoless double $\beta$-decay, which requires the $ee$ element of the 
${\cal M}_\nu$ matrix to be bounded as~\cite{Baudis}
\bea {\cal M}_{ee}<B=0.2\ {\rm eV}.
\label{B}
\eea
In addition, Tritium $\beta$-decay
experiments indicate $m_{\nu_i} < 2.5$ eV for any mass eigenstate with
a significant $\nu_e$ component \cite{triti}. Finally, concerning the
cosmological relevance of neutrinos, we will
take $\sum m_{\nu_i} = 6$ eV as a typical possibility and we will explain 
how the results vary when this value is changed.

\vspace{0.2cm}
\noindent
Let us introduce now some notation. In the SUSY framework the effective
mass term for the three light (left-handed) neutrinos in the flavour basis
is given by a term in the superpotential
\bea {W_{eff}}=\frac{1}{2} \nu^T {\cal M_\nu}  \nu\; .
\label{Mnu}
\eea
The mass matrix, ${\cal M_\nu}$, is diagonalized in the usual way, i.e.
${\cal M_\nu} = V^* D\, V^\dagger$, where $D=
{\rm diag}(m_1e^{i\phi},m_2e^{i\phi'},m_3)$ and
$V$ is a unitary `CKM' matrix,
relating  flavour to mass eigenstates
\bea 
\pmatrix{\nu_e \cr \nu_\mu\cr
\nu_\tau\cr}= \pmatrix{c_2c_3 &   c_2s_3 &   s_2e^{-i\delta}\cr
-c_1s_3-s_1s_2c_3e^{i\delta} &   c_1c_3-s_1s_2s_3e^{i\delta} &
s_1c_2\cr s_1s_3-c_1s_2c_3e^{i\delta} &   -s_1c_3-c_1s_2s_3e^{i\delta}
&   c_1c_2\cr}\, \pmatrix{\nu_1\cr \nu_2\cr \nu_3\cr}\,.
\label{CKM}
\eea 
Here $s_i$ and $c_i$ denote $\sin\theta_i$ and $\cos\theta_i$,
respectively. In the following we will label the mass eigenvectors
$\nu_i$ as $m_{\nu_1}^2<m_{\nu_2}^2$ and 
$|\Delta m^2_{12}|<|\Delta m^2_{23}|$, where $\Delta m^2_{ij}\equiv
m^2_{j}-m^2_{i}$ ($m_{\nu_3}^2$ is thus the most split eigenvalue).
In this notation, constraint (\ref{B}) reads \bea {\cal M}_{ee} \equiv
\big\vert  m_{\nu_1}\,c_2^2c_3^2e^{i\phi} +
m_{\nu_2}\,c_2^2s_3^2e^{i\phi'} + m_{\nu_3}\, s_2^2\,e^{i2\delta}
\big\vert < B \,.
\label{doublebeta}
\eea
As it has been put forward by Georgi and Glashow in ref.~\cite{GG}, a
scenario of nearly degenerate neutrinos should be close
to a bimaximal mixing, which constrains the texture of the mass matrix
${\cal M}_\nu$ to be essentially~\cite{GG,barger}  
\bea 
{\cal M}_{b} = m_\nu \,\pmatrix{
0 & {\displaystyle{1\over\sqrt2}} &  {\displaystyle{1\over\sqrt2}}\cr
{\displaystyle{1\over\sqrt2}} &
 {\displaystyle{1\over2}} &
- {\displaystyle{1\over2}}\cr  {\displaystyle{1\over\sqrt2}} &  -
{\displaystyle{1\over2}} &
 {\displaystyle{1\over2}}\cr }\;,
\label{MGG}
\eea where $m_\nu$ gives an overall mass scale.  ${\cal M}_{b}$
can be
diagonalized by a $V$ matrix  \bea {V}_{b} =  \,\pmatrix{
 {\displaystyle{-1\over\sqrt2}} &   {\displaystyle{1\over\sqrt2}} &  0 \cr
{\displaystyle{1\over 2}}
&  {\displaystyle{1\over2}} &
 {\displaystyle{-1\over\sqrt2}} \cr  {\displaystyle{1\over2}} &
{\displaystyle{1\over2}} &
   {\displaystyle{1\over\sqrt2}} \cr }\;,
\label{VGG}
\eea 
leading to exactly degenerate neutrinos, $D=m_\nu\  {\rm diag}(-1, 1,
1)$, and $\theta_2=0$, $\sin^2 2\theta_3=\sin^2 2\theta_1=1$.

It is quite conceivable that ${\cal M}_\nu={\cal M}_{b}$ could be
generated
at some high scale by interactions obeying appropriate continuous or
discrete symmetries \cite{sym}. However, in order to be
realistic, ${\cal M}_\nu$ at low energy should be {\em slightly}
different from ${\cal M}_b$ to account for the mass splittings
given in
eqs.(\ref{atm}--\ref{VO}). We will explore whether the 
appropriate splittings (and mixing angles) can
be generated or not through radiative corrections; more precisely,
through the running of the RGEs from the high scale down to low
energy. As discussed in ref.\cite{cein}, the output of this analysis 
can be of three types:

\begin{description}

\item[{\em i)}] All the mass splittings and mixing angles obtained
from the RG running are in agreement with all  experimental limits and
constraints.

\item[{\em ii)}] Some (or all) mass splittings are much larger than
the acceptable ranges.

\item[{\em iii)}] Some (or all) mass splittings are smaller than the
acceptable ranges, and the rest is within.

\end{description}

Case {\em (i)} is fine, while case {\em (ii)} is disastrous.
Case {\em (iii)} is not fine, but it admits the possibility that other 
(not specified) effects could be responsible for the splittings.
Concerning the mixing angles, it has been stressed in refs.~\cite{EL,cein}
that due to the
two degenerate eigenvalues of ${\cal M}_{b}$, ${V}_{b}$ is not
uniquely defined. Hence, once the ambiguity is
removed thanks to the small splittings coming from the RG running,
the mixing angles may be very different from the desired
ones. If such cases correspond to the previous {\em (iii)}
possibility, they could still be rescued since the modifications 
on ${\cal M}_\nu$ (of non-specified origin) needed to reproduce the correct
mass
splittings will also change dramatically the mixing angles.

In section 2, we examine the general case in which the neutrino
masses arise from an effective operator, remnant from new physics entering
at a scale $\Lambda$. In this framework, we assume a bimaximal-mixing mass
structure at the scale
$\Lambda$ as an initial condition and do not consider
possible
perturbations of that initial condition coming from the new physics
entering at $\Lambda$.
If $\tan\beta$ is small the LAMSW scenario in this case is of the
undecidable type [possibility {\em (iii)} above], but for $\tan\beta$ above
a certain value which we compute, it is excluded (the
VO solution is excluded for any $\tan\beta$).

In section 3 we consider in detail a particularly
well motivated example for the new physics beyond the scale $\Lambda$
introduced before: the see-saw scenario. We include
here the high energy effects of the new degrees of freedom above the scale
$\Lambda$ (identified now with the mass scale of the right-handed
neutrinos).
We find regions of parameter space
where the neutrino spectrum and mixing angles fall naturally in the
pattern required to explain solar (LAMSW solution) and atmospheric
neutrino
anomalies, which we find remarkable. We complement the numerical results,
presented in section 4,
with compact analytical formulas which give a good description of them,
and
allow to understand the pattern of mass splittings and mixing angles
obtained. 
We also present  plausible textures for the neutrino Yukawa couplings
leading
to a good fit of the oscillation data. 

Section 5, still in the see-saw framework, discusses several possible
implications of the effect of large neutrino couplings on:
triviality limits on the Majorana mass; 
the infrared fixed point value of the top Yukawa coupling, with
consequences for the lower limit on $\tan\beta$ and the value of the Higgs
mass; and gauge coupling and bottom-tau unification.
Finally we draw some conclusions.

\section{${\cal M}_\nu$ as an effective operator}

In this section we will simply assume that the effective mass matrix
for the left-handed neutrinos, ${\cal M}_\nu$, is generated at some
high energy scale, $\Lambda$, by some unspecified mechanism. Assuming
that below $\Lambda$ the effective theory is the MSSM with unbroken
$R-$parity, the lowest dimension operator in the superpotential
producing a mass of this kind is \cite{Babu}
\bea {W_{eff}}=\frac{1}{4}\kappa \nu^T   \nu H_2^0 H_2^0 \; ,
\label{kappa}
\eea
where $\kappa$ is a matricial coupling and $H_2^0$ is the neutral component 
of
the 
$Y=+1/2$ Higgs field (the one coupled to the $u-$quarks). Obviously,
${\cal
M}_\nu=\frac{1}{2}\kappa \langle H_2^0\rangle^2$. Between $\Lambda$ and 
$M_{SUSY}$, the effective coupling
$\kappa$ runs with the scale with a RGE \cite{Babu}
\bea 
16\pi^2 \frac{d \kappa}{dt}= \kappa
\left[-\frac{6}{5}g_1^2-6g_2^2+6Y_t^2\right]
+\left[\kappa{\bf
Y_e^\dagger Y_e} + ({\bf Y_e^\dagger Y_e})^T\kappa\right]\ ,
\label{rg1}
\eea
where $t=\log \mu$, and  $g_2, g_1, Y_t, {\bf Y_e}$ are the
$SU(2)\times U(1)_Y$ gauge couplings, the (MSSM) top Yukawa
coupling and the (MSSM) matrix of Yukawa couplings for the 
charged leptons respectively.
Between $M_{SUSY}$ and $M_Z$ $\kappa$ runs with the SM RGE \cite{Babu}
%
%
\bea 
16\pi^2 \frac{d \kappa}{dt}= \left[-3g_2^2+2\lambda+6{\tilde{Y_t}}^2+2
{\rm
Tr}{\bf \tilde{Y}_e^\dagger \tilde{Y}_e} \right]\kappa
-\frac{1}{2}\left[\kappa{\bf
\tilde{Y}_e^\dagger \tilde{Y}_e} + ({\bf \tilde{Y}_e^\dagger
\tilde{Y}_e})^T\kappa\right]\ ,
\label{rg1SM}
\eea 
where $\lambda$ is the SM quartic Higgs coupling and $\tilde{Y_t}$,
${\bf \tilde{Y}_e}$ correspond to the SM Yukawa couplings [the matching at
$M_{SUSY}$ is
$\tilde{Y_t}(M_{SUSY})=\sin\beta \, {Y_t}(M_{SUSY})$,
${\bf \tilde{Y}_e}(M_{SUSY})=\cos\beta\, {\bf Y_e}(M_{SUSY})$].

In a scenario of almost degenerate neutrinos, the simplest assumption
for the initial form of the matricial coupling, 
$\kappa(\Lambda)$ is just
the bimaximal mixing texture of eq.(\ref{MGG}),
and this was also the assumption made in refs.\cite{EL,cein}. In
consequence,
\bea {\cal M}_\nu(\Lambda)=\frac{1}{2}\kappa(\Lambda) \langle
H_2^0\rangle^2=\frac{1}{2}\kappa(\Lambda) \sin^2\beta v^2=
{\cal M}_{b}\, ,
\label{kappaboundary}
\eea
where $v^2=\langle H_1^0\rangle^2+\langle H_2^0\rangle^2=(175\, {\rm
GeV})^2$. The last terms 
in  eqs.(\ref{rg1}, \ref{rg1SM}), i.e. those depending on ${\bf 
Y_e^\dagger Y_e}$, are generation-dependent and will modify the ${\cal
M}_{\nu}$ texture, thus
generating mass
splittings and changing the mixing angles. It is easy to see
\cite{EL,cein}
that, in  first approximation, the splittings $\Delta m^2_{ij}\equiv
m^2_{j}-m^2_{i}$
have the form
\bea 
\Delta m^2_{12}\ =\ -\frac{1}{2}\Delta m^2_{13}\ =\
-\frac{1}{3}\Delta m^2_{23} \ \simeq\ - m_\nu^2 {\epsilon}>0.
\label{Deltaskappa}
\eea
%
%
%
where, neglecting for a while all the charged lepton Yukawa
couplings but $Y_\tau$, and working in the approximation of constant
RGE $\beta$ functions, $\epsilon$ is given by
\bea
\epsilon =\frac{\tilde{Y_\tau}^2}{32\pi^2}\left[
-\frac{2}{\cos^2\beta}\log\frac{\Lambda}{M_{SUSY}}
+\log\frac{M_{SUSY}}{\mu_0}\right]\, .
\label{eps}
\eea
Thus, the SUSY and the SM corrections have opposite signs. If
$({\Lambda}/M_{SUSY})^{2/\cos^2\beta}>{M_{SUSY}}/{\mu_0}$, as it is
the usual case, $\epsilon$ has negative sign and the most split
eigenvalue is the smallest one [thus the convention of labels used in
eq.(\ref{Deltaskappa})].  Notice also that the pure SM case is
recovered setting  $M_{SUSY}=\Lambda$. As in the SM case, the previous
spectrum is not realistic, i.e. the splittings are barely able to
reproduce simultaneously the $\Delta m^2_{sol}$ and $\Delta m^2_{at}$
splittings given in eqs.(\ref{atm}--\ref{VO}). 
Actually, the size of the splittings is larger
than in the pure SM case (due to the coefficients of the RGEs and,
especially, to the dependence on $\cos\beta$). This makes the scenario
potentially more difficult than the SM one (see the discussion in the
Introduction).
%
\begin{figure}[t]  
\centerline{
\psfig{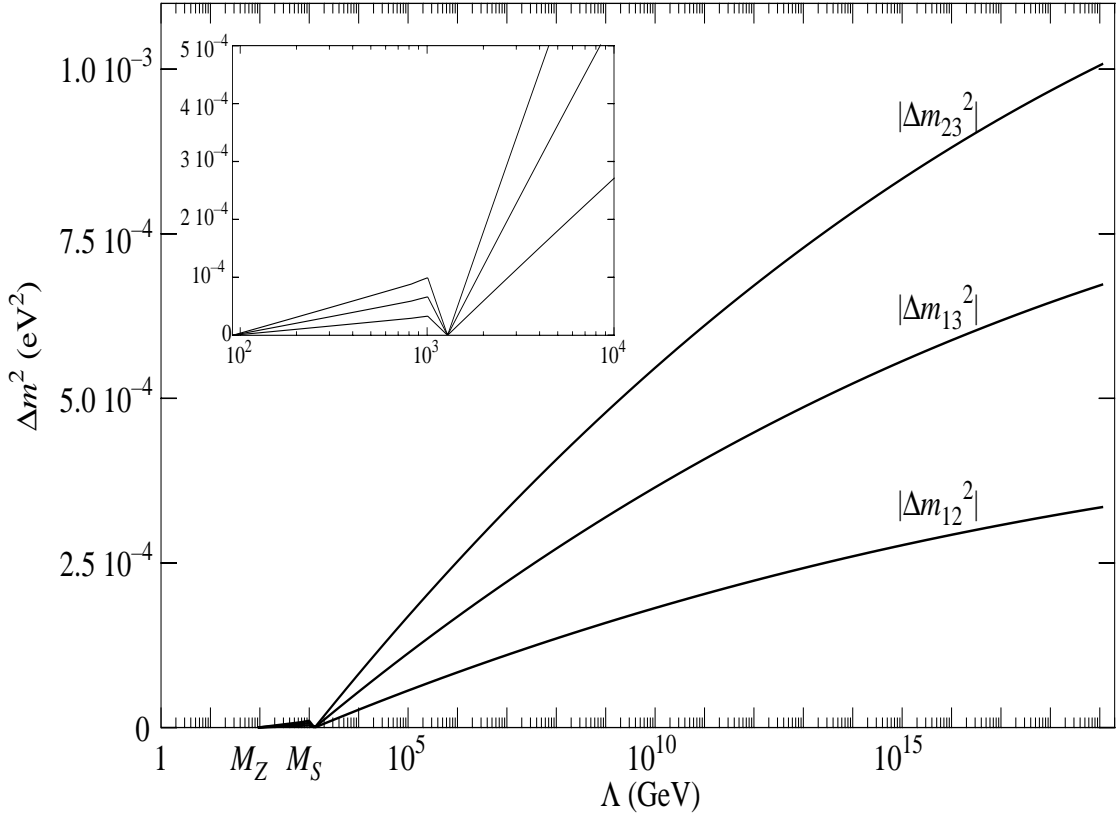}}
\caption
{\footnotesize 
Dependence of neutrino mass splittings at low energy ($\Delta m_{ij}^2$ in
${\mathrm eV}^2$) with the cut-off scale $\Lambda ({\rm GeV}$). For this
figure $\tan\beta=2$.  
}
\end{figure}

Fig.1 shows the complete numerical evaluation of the RGEs for
$m_\nu=2$ eV and $\tan\beta=2$, which corroborates the structure of
eqs.(\ref{Deltaskappa}, \ref{eps}). The splittings
are always much larger than those required for the VO solution to the
solar neutrino problem, $\Delta m^2_{sol}\sim
10^{-10}$ eV$^2$. Therefore, the effect of the RGEs for
this scenario is disastrous in the sense discussed in the Introduction
for the possibility {\em (ii)}. In consequence, as for the pure SM case, 
the VO solution to the solar neutrino problem is excluded\footnote{
Notice from the figure and from eq.(\ref{eps})
that there is a value of $\Lambda$ (close to $M_{SUSY}$) for which the
splittings vanish, but of course the fine-tuning required for $\Lambda$ 
to be close enough to that point, so that the splittings are within the
VO range, is enormous.}.
For the LAMSW solution to the solar neutrino problem, things are a bit
different. The smallest mass splitting is always within 
(or not far larger than) the LAMSW
range, which indicates that we are in the situation $(iii)$ explained in
the Introduction
(not satisfactory, but it may be rescued by extra physics).
When $m_\nu$ is varied the results change with the scaling law
$\Delta m_{ij}^2\propto m_\nu^2$.
%
\begin{figure}[t]  
\centerline{
\psfig{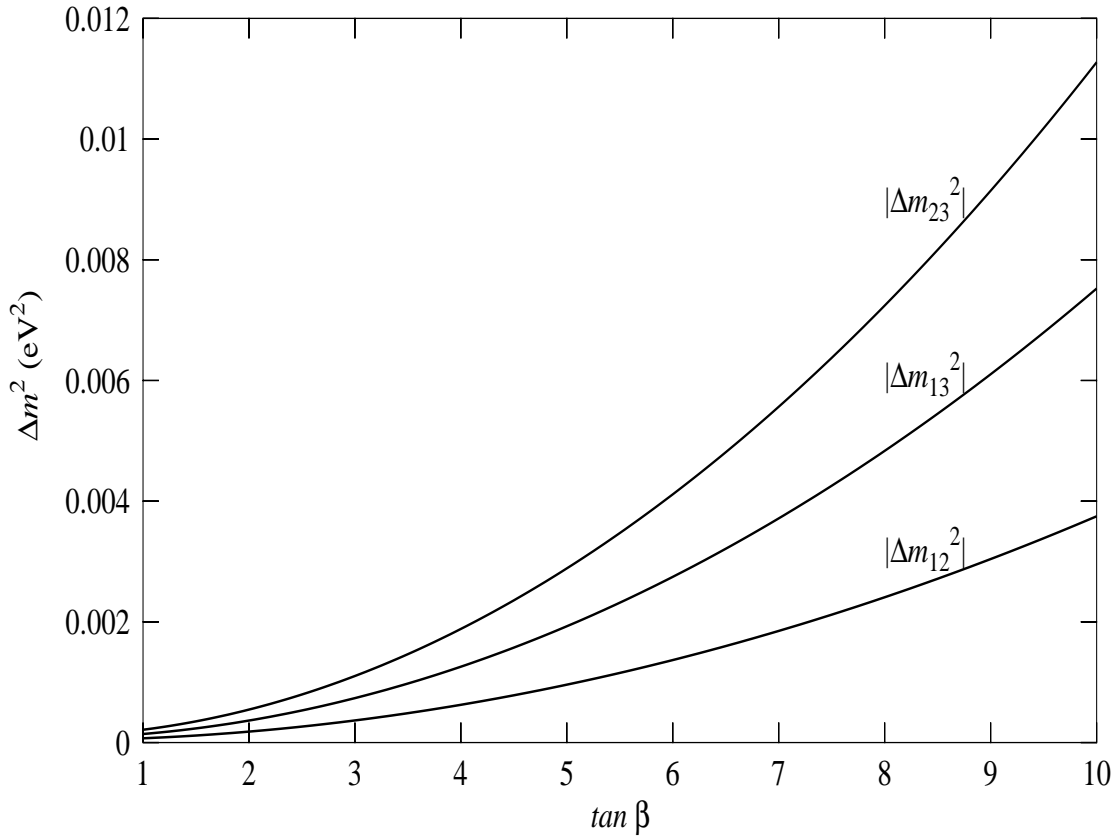}}
\caption
{\footnotesize Dependence of neutrino mass
splittings at
low energy ($\Delta m_{ij}^2$ in ${\mathrm eV}^2$) with $\tan\beta$,
for $\Lambda=10^{10} {\rm GeV}$.
}
\end{figure}

The dependence of the
splittings on $\tan\beta$ is quite strong. Fig.2 shows this
dependence for $\Lambda=10^{10}$ GeV. 
Even for moderate values of
$\tan\beta$ the smallest splitting is much larger than the 
LAMSW range, thus spoiling that solution. Therefore, for a given
value of $\Lambda$ and $m_\nu$, the viability of the supersymmetric
scenario of
nearly degenerate neutrinos puts an upper bound on $\tan \beta$.
A reasonable estimate for this upper bound is given in Fig.3, which shows 
the value of $\tan\beta$ (vs. $\Lambda$) for which the smallest splitting
generated is larger (and ten times larger), than the maximum acceptable
LAMSW value.
The upper curve is therefore a conservative bound.
These bounds become weaker if the neutrino mass, $m_\nu$, decreases,
as it follows from the previously mentioned scaling law.
%
\begin{figure}[t]  
\centerline{
\psfig{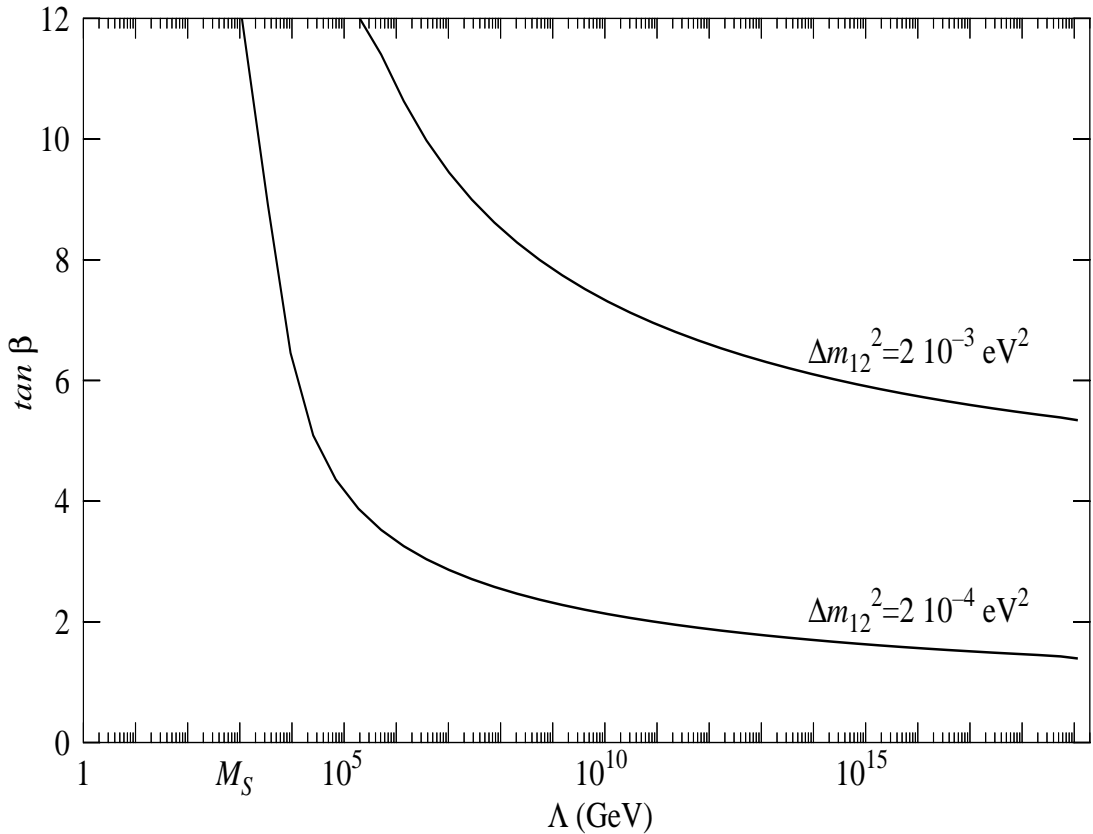}}
\caption
{\footnotesize Upper limit on $\tan\beta$, as a function of $\Lambda$,
beyond which the smallest neutrino mass splitting is larger (lower curve),
or ten times larger (upper curve), than the maximum acceptable LAMSW
value.
}
\end{figure}

Concerning the mixing angles, in a first approximation the 
`CKM' matrix, $V$, is given by
\bea V\simeq\pmatrix{  
-{\displaystyle{1\over\sqrt2}} &
{\displaystyle{1\over\sqrt3}} &   
 {\displaystyle{1\over\sqrt6}}\cr  
 {\displaystyle{1\over2}} &
{\displaystyle{\sqrt{2\over3}}} &  
- {\displaystyle{1\over2\sqrt3}}\cr  
{\displaystyle{1\over2}} &  
0 &  
{\displaystyle {\sqrt3\over2}}\cr
}\, ,
\label{Vkappa}
\eea
which leads to mixing angles
\bea \sin^2 2 \theta_1 = {9\over25},\;\;\; \sin^2 2 \theta_2 =
{5\over9},\;\;\; \sin^2 2 \theta_3 =  {24\over25}.
\label{mixingsskappa}
\eea

These values are far away from the bimaximal mixing ones.
In consequence, they are not acceptable, stressing the fact that the
simplified scenario discussed in this section does not work.
However, as explained in the Introduction, if extra effects are able to 
modify the form of ${\cal M}_\nu$ in order to produce the correct
splittings, this modification will also change drastically the mixing
angles, hopefully in a positive direction. This leads us to the see-saw
scenario, which is analyzed in the next section.

\section{${\cal M}_\nu$ from the see-saw mechanism}

The simplest example of the kind of new physics appearing at a scale
$\Lambda$ which can generate an effective mass term for the low-energy
neutrinos we observe is the so-called see-saw mechanism \cite{seesaw}.
Its supersymmetric version has superpotential
\bea
\label{superp}
W=W_{MSSM} - \frac{1}{2}\nu_R^c{\cal M}\nu_R^c
+\nu_R^c {\bf Y_\nu} L\cdot H_2,
\eea
where $W_{MSSM}$ is the superpotential of the MSSM. The extra terms 
involve three additional neutrino chiral fields (one per generation;
indices are suppressed) not charged under the SM group: $\nu_{\alpha,R}$
($\alpha=e,\mu,\tau$). 
${\bf Y_\nu}$ is the matrix
of neutrino Yukawa couplings and $H_2$ is the hypercharge $+1/2$ Higgs
doublet. Now, the Dirac mass matrix is ${\bf m_D}={\bf Y_\nu}v \sin\beta$. 
Finally, ${\cal M}$ is a $3\times 3$ Majorana mass matrix which does not
break the SM gauge symmetry. It is natural to assume that the
overall scale of ${\cal M}$, which we will denote  by $M$, is much larger
than the electroweak scale or any soft mass. Below $M$ the theory is 
governed by an effective superpotential
\bea
W_{eff}=W_{MSSM}+\frac{1}{2}({\bf
Y_\nu}L\cdot H_2)^T{\cal M}^{-1}({\bf
Y_\nu}L\cdot H_2),
\eea
obtained by integrating out the heavy neutrino fields in (\ref{superp}).
From this effective superpotential, the Lagrangian contains a mass term
for the left-handed neutrinos:
\bea
\delta {\cal L}=-\frac{1}{2}\nu^T{\cal M}_\nu \nu + {\rm h.c.},
\eea
with
\bea {\cal M}_\nu=
{\bf m_D}^T {\cal M}^{-1} {\bf m_D} =  {\bf Y_\nu}^T {\cal M}^{-1} {\bf
Y_\nu}
\langle H_2^0\rangle^2, 
\eea 
suppressed with respect to the typical
fermion masses by the  inverse power of the large scale $M$.

The approximate degeneracy of neutrino masses in this framework follows
from the initial condition for the neutrino mass matrix at
the Planck scale: ${\cal M}_\nu(M_p)={\cal M}_{b}$, which would lead to
exactly degenerate neutrino masses. We do not address in this paper the
possible origin of this very symmetric form. 
 As explained in \cite{cein}, to
explore the simplest (and most natural) textures of ${\bf Y_\nu}$
and ${\cal M}$ leading to this
initial condition it is enough to consider the case in which the Majorana
mass matrix is
\bea
{\cal M}=M\pmatrix{-1 & 0 & 0\cr
0 & 1 & 0\cr
0& 0& 1}\, ,
\label{Mmaj}
\eea
and all the structure  is in the Yukawa matrix, which reads
\bea 
{\bf Y_\nu}= Y_\nu B V_{b}^T \, .
\label{Ynu}
\eea   
Here $Y_\nu$ is the overall magnitude of ${\bf Y_\nu}$ and $B$ is a
combination of two `boosts'
\bea
B=\pmatrix{ \cosh a & 0 &  \sinh a
\cr 0 &  1 &  0 \cr  \sinh a   &     0    & \cosh a  \cr} \pmatrix{
\cosh b   &  \sinh b   &   0     \cr  \sinh b  &  \cosh b   & 0 \cr 0
   &   0  & 1 \cr} \, ,
\label{boosts}
\eea
with two free parameters $a,b$. Having all the structure in the Majorana
matrix and ${\bf Y_\nu}\propto {\bf I_3}$ is equivalent to the case
$a=b=0$. We will present our results, for fixed values of $M$ and
$\tan\beta$, in the $(a,b)$ plane demanding $|a|,|b|\leq 1.5$. If $|a|$ or
$|b|$ are larger than 1.5 the matrix elements of ${\bf Y_\nu}$ are
fine-tuned at least in a $10\%$ \cite{cein}. 

The neutrino masses, exactly degenerate at tree-level by assumption, will
receive generation dependent radiative corrections which will lift that
degeneracy, eventually reproducing the pattern of mass splittings
necessary to interpret the experimental indications. The bulk of these
radiative corrections is logarithmic and easy to compute by
standard renormalization group techniques: one starts at $M_p$ with 
(\ref{Mmaj}) and (\ref{Ynu}) as boundary conditions and integrates down
in energy the relevant RGEs in a supersymmetric theory which is the MSSM
with three right-handed neutrino chiral fields and the superpotential
(\ref{superp}). At the scale $M$ the $\nu_{R,\alpha}$ are decoupled and
below this scale the running is performed exactly as in the previous
section, except that now we have precise boundary conditions for the
couplings and masses at $M$.

From $M_p$ to $M$ the evolution of the relevant matrices is
governed by the following renormalization group equations~\cite{RGE}: 
\bea
\label{rg2}
\frac{d {\bf Y_\nu}}{dt}= -\frac{1}{16\pi^2} {\bf Y_\nu}\left[\left(
3 g_2^2+\frac{3}{5}g_1^2-{\mathrm T_2}
\right){\bf I_3}-\left( 3 {\bf
Y_\nu}^\dagger  {\bf Y_\nu}+{\bf Y_e^\dagger Y_e}\right) \right], \eea
\bea
\label{rg3}
\frac{d {\bf Y_e}}{dt}= -\frac{1}{16\pi^2} {\bf Y_e} \left[\left(
3 g_2^2+\frac{9}{5}g_1^2-{\mathrm T_1}
\right){\bf I_3}-\left(
{\bf Y_\nu}^\dagger {\bf Y_\nu}+ 3  {\bf Y_e^\dagger Y_e}\right) \right],
\eea 
where
\bea
{\mathrm T_1}={\mathrm Tr}(
3 {\bf Y_D^\dagger Y_D}+{\bf Y_e^\dagger Y_e}),\;\;\;
{\mathrm T_2}={\mathrm Tr}(
3{\bf Y_U^\dagger Y_U}+{\bf Y_\nu^\dagger Y_\nu}),
\eea
and
\bea
\label{rg4}
\frac{d {\cal M}}{dt}=\frac{1}{8\pi^2}\left[{\cal M} ({\bf Y_\nu}
{\bf Y_\nu}^\dagger)^T+{\bf Y_\nu} {\bf Y_\nu}^\dagger {\cal M}\right],
\eea 
(not yet given in the literature). Here $g_2$ and $g_1$ are the
$SU(2)_L$ and $U(1)_Y$ gauge coupling constants, and ${\bf Y_{U,D,e}}$ are
the Yukawa matrices for up quarks, down quarks and charged leptons.

At $M$, $\nu_R$ decouple, and ${\bf Y_e}$ must be diagonalized to
redefine the flavour basis of leptons [note that the last term in
(\ref{rg3}) produces non-diagonal contributions to ${\bf Y_e}$] affecting the
form of the ${\bf Y_\nu}$ matrix. 
Then the effective mass matrix for
the light neutrinos is ${\cal M}_\nu\simeq {\bf Y_\nu}^T {\cal M}^{-1} {\bf
Y_\nu} \langle H_2^0\rangle^2$.

From $M$ to $M_Z$, the effective mass matrix ${\cal M}_\nu$ is
run down in energy exactly as described in section 2.

The renormalization group equations are
integrated with the following boundary conditions:
${\cal M}$ and ${\bf Y_\nu}$ are chosen at $M_p$ so as to satisfy \bea
{\cal M}_\nu(M_p)={\cal M}_{b}, \eea with the overall magnitude of
${\bf Y_\nu}$ fixed, for a given value of the Majorana mass $M$, by the
requirement $m_{\nu}\sim {\cal O}({\rm eV})$. We will take $m_\nu=2$ eV as
a guiding example. The boundary conditions for
the other Yukawa couplings are also fixed at the low energy  side to give
the observed fermion masses.  The free parameters are therefore $M,
\tan\beta, a$ and $b$.

\subsection{Analytical integration of the RGEs}

It is simple and very illuminating to integrate analytically
the renormalization group equations
(\ref{rg1},\ref{rg2},\ref{rg3},\ref{rg4}) 
in the approximation of constant right hand side. In this
approximation (which works very well for our analysis), the effective
neutrino mass matrix at low-energy is simply ${\cal M}_{b}$ plus some small
perturbation. The overall mass scale is fixed to be of order 2 eV, so we
need only to pay attention to the non-universal terms in the RGEs, which
will be responsible for the mass splittings.
Neglecting the $Y_e$, $Y_\mu$
Yukawa couplings, we get the following analytical
expressions (the labelling of mass eigenvalues below may
not always correspond to the conventional order
$m_{\nu_1}^2<m_{\nu_2}^2$, $|\Delta m_{12}^2| <|\Delta m^2_{23}|$):
\bear{cl}
m_{\nu_1}\simeq &m_\nu \left[-1+
(2 c_a^2 c^2_b-1)\epsilon_\nu-2\epsilon_\tau
\right],\vspace{0.2cm}\\
m_{\nu_{2,3}}\simeq &m_\nu \left[1+3 \epsilon_\tau - c^2_a c^2_b
\epsilon_\nu\pm \left\{\left[\epsilon_\tau+(c^2_a c^2_b-c_{2a})\epsilon_\nu
\right]^2 +\left[s_{2a} s_b \epsilon_\nu - 2\sqrt{2}\epsilon_\tau 
\right]^2
\right\}^{1/2} \right]
\label{mnu123}
\eear
where $c_a=\cosh a$, $s_{2a}=\sinh 2a$, etc. These expressions are
identical to
the ones derived in \cite{cein} for the SM case, except for the numerical
values of $\epsilon_\tau$ and $\epsilon_\nu$, which are now given by:
\bea
\label{epstau}
\epsilon_\tau =\frac{Y_\tau^2}{128\pi^2}\left[\
-2\log\frac{M_p}{M_{SUSY}}+\cos^2\beta\log\frac{M_{SUSY}}{M_Z}\right],
\eea
\bea
\label{epsnu}
\epsilon_\nu=\frac{Y_\nu^2}{8\pi^2}\log\frac{M_p}{M}.
\eea
The Yukawa couplings in these expressions should be chosen at an
appropriate intermediate scale but the simple formulas
(\ref{epstau},\ref{epsnu}) are good enough for our purpose. Note also that 
$Y_\nu$ and $Y_\tau$ have an implicit dependence on $\tan\beta$ and,
furthermore $Y_\nu$ depends strongly on $M$ (due to the requirement
$m_\nu\sim 2$ eV).

There are important differences with respect to the SM case presented in
\cite{cein}: 1) $\epsilon_\tau$ is insensitive to the Majorana threshold,
 its sign is negative (it was positive in the SM) and it grows in
magnitude for increasing $\tan\beta$; 2) $\epsilon_\nu$ is
twice larger than in the SM and decreases slightly when
$\tan\beta$ increases.

In the case $a=b=0$, the mass splittings are independent of
$\epsilon_\nu$ and similar to those found in section 2, that is, not
satisfactory (remember that this case is equivalent to having all
the structure in ${\cal M}$, while
${\bf Y}_{\nu}$ is proportional to the identity and thus universal). 

For a given $M$, there is a critical value of $\tan\beta$ below
(above) which $|\epsilon_\tau|$ is smaller (larger) than $\epsilon_\nu$.
This critical value of $\tan\beta$ increases with $M$.
When $\epsilon_\nu\gg |\epsilon_\tau|$, we can further expand the neutrino
masses in powers of $\epsilon_\tau/\epsilon_\nu$ finding
\bear{cl}
m_{\nu_1}\simeq &m_\nu \left[-1+
(2 c_a^2 c^2_b-1)\epsilon_\nu-2\epsilon_\tau
\right],\vspace{0.2cm}\\
m_{\nu_2}\simeq &m_\nu
\left[1-(2 c_a^2
c^2_b-1)\epsilon_\nu+\left(2-{\displaystyle\frac{1-c_{2a}-2\sqrt{2}
s_{2a}s_b}{c_a^2
c^2_b-1}}
\right)\epsilon_\tau\right]
,\vspace{0.2cm}\\
m_{\nu_3}\simeq &m_\nu
\left[1-\epsilon_\nu+\left(4+{\displaystyle\frac{1-c_{2a}-2\sqrt{2}
s_{2a}s_b}{c_a^2
c^2_b-1}}
\right)\epsilon_\tau\right].
\label{mnu1232}
\eear
Here we clearly see that the small mass splitting (solar)
is controlled by the small parameter
 $\epsilon_\tau$, proportional to the squared Yukawa couplings of the charged
leptons, while $\epsilon_\nu$ (proportional to the square of 
the larger Yukawa coupling $Y_\nu$) is responsible for the 
larger mass difference (atmospheric). Moreover, the two neutrino
eigenstates closest in mass ($m_{\nu_1}, m_{\nu_2}$) have masses of
opposite sign, which is exactly 
what is required to fulfill the neutrinoless double $\beta$-decay
condition, eq.(\ref{doublebeta}). In the SM case, for values of $M$
between $10^9$ and $10^{12}$ GeV, $\epsilon_\tau$ and $\epsilon_\nu$ have
the right orders of magnitude to account for $\Delta m^2_{sol}$, $\Delta
m^2_{at}$ \cite{cein}. In the supersymmetric scenario it is still the case
that $\epsilon_\nu$ gives the correct atmospheric mass splitting, but
for $m_\nu=2$ eV
$\epsilon_\tau$ gives typically a $\Delta m^2_{sol}$ one order of
magnitude too high (recall here that by lowering $m_\nu$ this 
splitting becomes smaller following the approximate law
$\Delta m^2\propto m_\nu^2$).
This means that in this case only in particular regions of the
$(a,b)$ plane, in which some mild cancellation is taking place, we will
obtain 
the right mass splitting to solve the solar neutrino problem. This
cancellation does occur in regions around the lines $a=0$ and
$2\sqrt{2}\sinh
b=-\tanh a$ for which, in the approximation (\ref{mnu1232}), $m_{\nu_1}^2=
m_{\nu_2}^2$. 
In those regions then, we have a natural 
explanation for the $\Delta m_{sol}^2$ mass splitting while $\Delta
m_{at}^2$ requires only a mild fine-tuning of parameters.

If $\epsilon_\nu\ll |\epsilon_\tau|$, then a different
expansion shows that there is no natural hierarchy of mass splittings,
which tend to be of the same order. In these cases, a stronger fine-tuning
is required to get small enough $\Delta m_{sol}^2$ and the regions in
parameter space where this occurs shrink, again around
$1-c_{2a}-2\sqrt{2} s_{2a}s_b=0$. However, $\Delta m_{at}^2$ is still
naturally of the right order of magnitude.

Turning to the mixing angles, it is easy to see that the eigenvectors
of the perturbed ${\cal M}_\nu$ matrix are of the form
\bea
V_1' = V_1,\;\;\;V_2' = \frac{1}{\sqrt{\alpha^2+\beta^2}}
(\alpha V_2+\beta V_3),\;\;\;
V_3' = \frac{1}{\sqrt{\alpha^2+\beta^2}}(-\beta V_2+\alpha V_3),
\label{eigenv}
\eea
where $V_i$ are the eigenstates corresponding to the bimaximal mixing
matrix,  $V_b$
[see eq.(\ref{VGG})]
\bea
V_1 = \left (   
\begin{array}{c}
{\displaystyle \frac{-1}{\sqrt{2}}} \vspace{.1cm}\\
{\displaystyle\frac{1}{2}}  \vspace{.1cm}\\
{\displaystyle\frac{1}{2}} 
\end{array}  
\right ), ~~    
V_2 = \left (   
\begin{array}{c}
{\displaystyle{1 \over \sqrt{2}}} \vspace{.1cm}\\ {\displaystyle{1 \over
2}}  \vspace{.1cm}\\
{\displaystyle{1 \over 2}}
\end{array}  
\right ), ~~    
V_3 = \left (   
\begin{array}{c}
0  \vspace{.1cm}\\ {\displaystyle\frac{-1}{\sqrt{2}}}  \vspace{.1cm}\\
{\displaystyle\frac{1}{\sqrt{2}}}
\end{array}
\right ).
\label{vecbim}
\eea
In the approximation $|\epsilon_\tau|\ll \epsilon_\nu$, the parameters 
$\alpha, \beta$ are given by
\bea
\alpha=c_{a}s_b +
{\cal O}({\epsilon_\tau}/{\epsilon_\nu}),\;\;\;\;\beta=s_{a}
+ {\cal O}({\epsilon_\tau}/{\epsilon_\nu}).
\label{alfabeta}
\eea
The $V_i'$ vectors define the new `CKM' matrix $V'$ from which the 
mixing angles are extracted. In the case
of interest, the spectrum of neutrinos consists of two lightest states 
($m_{\nu_{1,2}}$) very close in mass and a heavier one $m_{\nu_3}$.
The relative ordering in mass of $m_{\nu_{1,2}}$ is not fixed and
this affects the relative order of $V_1'$ and $V_2'$ in the $V'$ matrix.
This has an effect on the sign of $\cos 2\theta_3$, which is important for
the MSW condition
$\cos2\theta_{3}>0$ (written using the conventional 
order $m_{\nu_1}^2<m_{\nu_2}^2$). This will be  satisfied as long
as $V_1$ corresponds to the lightest mass eigenvalue. In other words,
this condition requires that
the negative mass eigenvalue, see eqs.(\ref{mnu123},
\ref{mnu1232}),
corresponds to the lightest neutrino. On the other hand, the relative
ordering of the two lightest neutrinos does not change the values of
$\sin^22\theta_i$.

In this approximation,  if   
just one of the two $(a,b)$ parameters is vanishing, then $V'=V_b$, i.e.
exactly the bimaximal mixing case.
Also, whenever $c_a$, $c_b$ are sizeable (i.e. away from $a=b=0$),
$|\alpha|\gg |\beta|$, and thus we are close to the bimaximal case.
Therefore, it is not surprising that in most of the parameter space 
this will be in fact the case. This is
remarkable, because it gives a natural origin for the bimaximal
mixing, which was not guaranteed a priori due to the ambiguity 
in the diagonalization of the
initial ${\cal M}_{\nu}(M_p)={\cal M}_{b}$ matrix, as was explained in
the Introduction.

The only free parameter in the $V'$ matrix is the
ratio $\alpha/\beta$ and one obtains the relations:
\bea
\sin^22\theta_1=\frac{(2r-1)^2}{(2r+1)^2},\;\;\;
\sin^22\theta_2=1-\frac{r^2}{(1+r)^2},\;\;\;
\sin^22\theta_3=1-\frac{1}{(1+2r)^2},
\label{tetas}
\eea
where $r\equiv \alpha^2/\beta^2$.
It is instructive to invert the last relation, find $r$ in terms of
$\sin^22\theta_3$, and then substitute back in the two previous relations. 

\begin{figure}[t]
\centerline{\vbox{
\psfig{figure=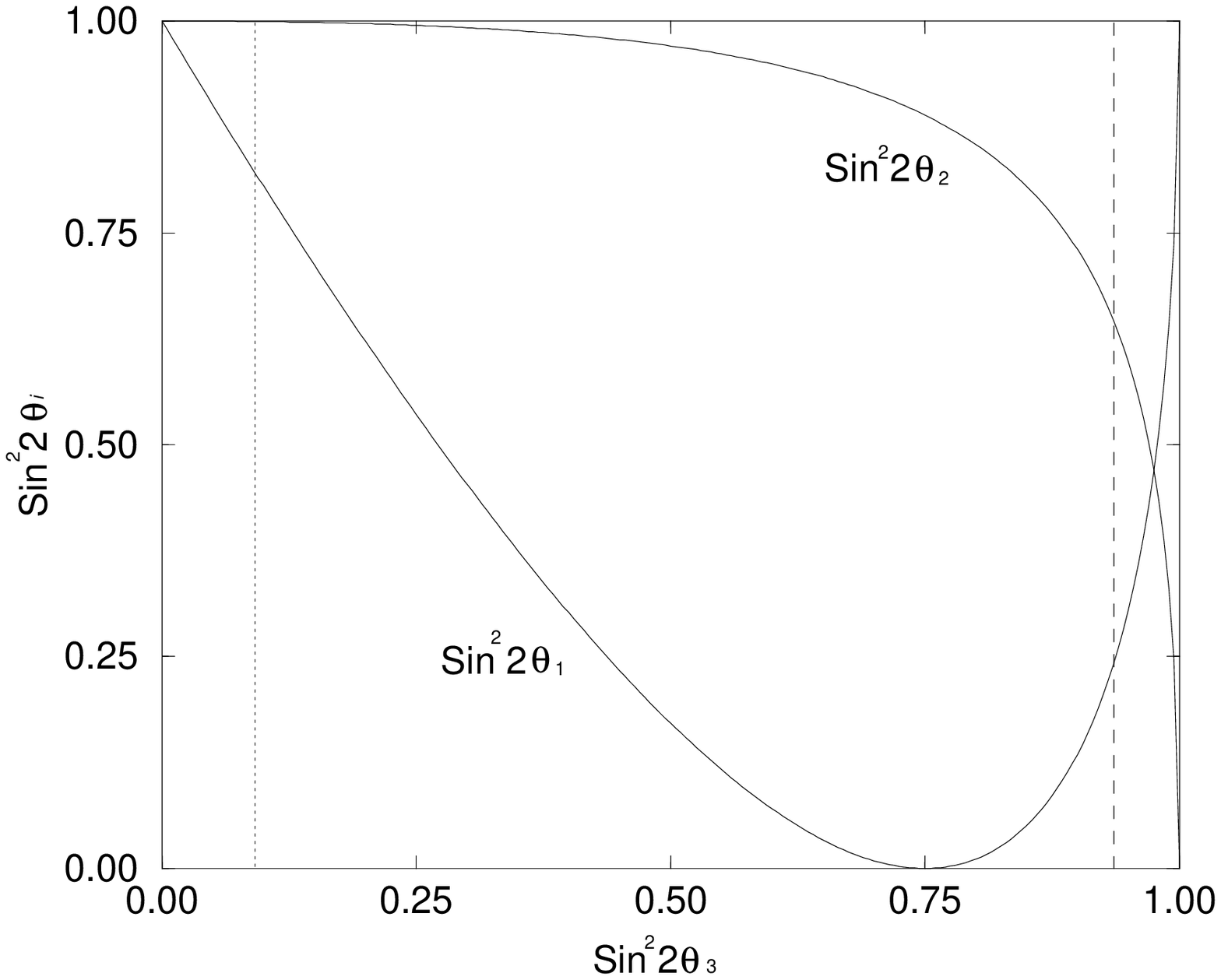,height=7cm,width=10cm,bbllx=4.cm,%
bblly=0.cm,bburx=17.cm,bbury=15.cm}
\psfig{figure=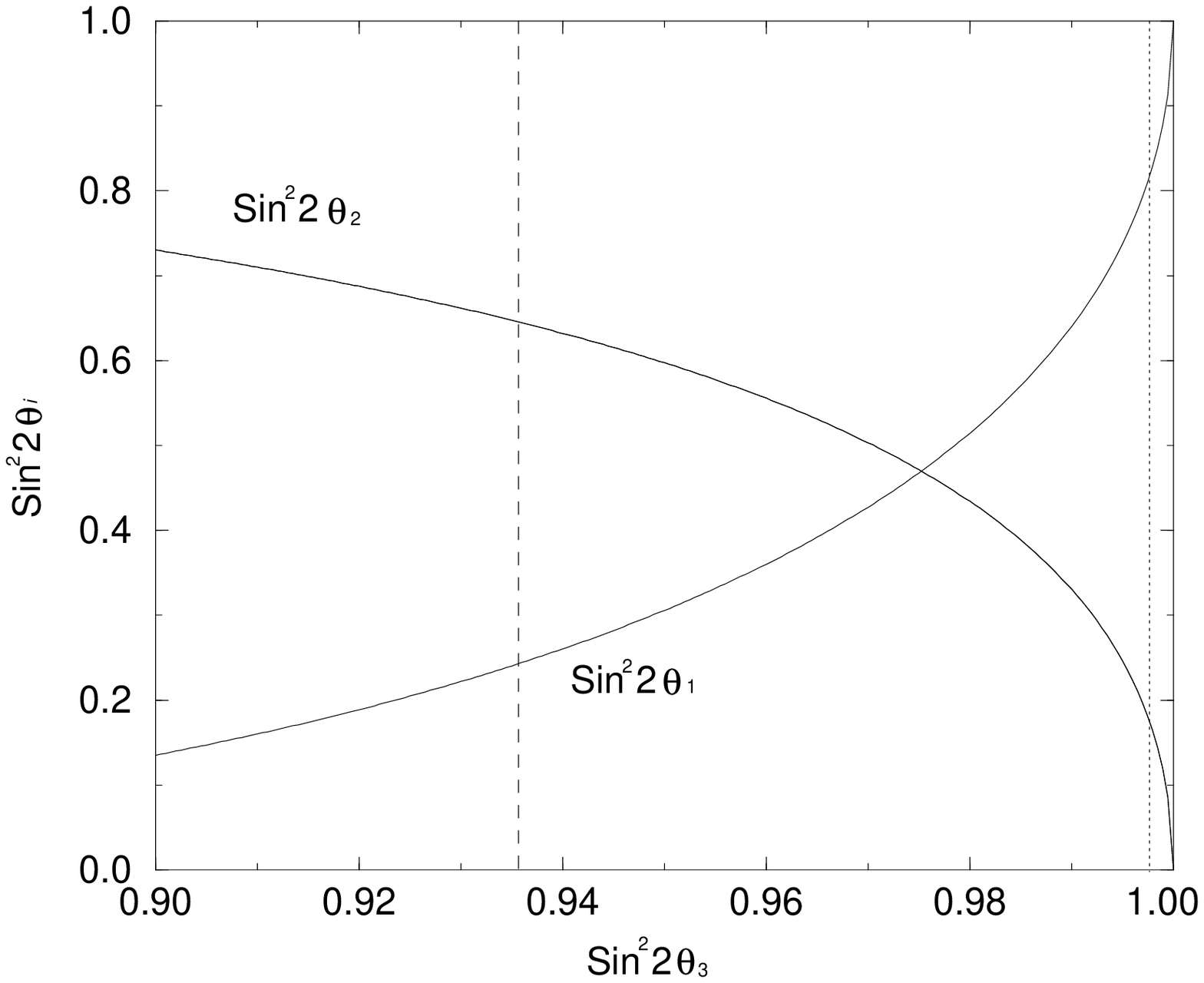,height=7cm,width=10cm,bbllx=4.cm,%
bblly=0.cm,bburx=17.cm,bbury=15.cm}}}
\caption{\footnotesize 
Upper plot: Neutrino mixing angles, $\sin^22\theta_1$ and
$\sin^22\theta_2$, vs.
$\sin^22\theta_3$ in the near degenerate scenario. 
Lower plot: Zoom of the $0.9\leq \sin^22\theta_3\leq 1$ region.
}
\end{figure}

Figures 4a,b present $\sin^22\theta_{1,2}$ as functions of
$\sin^22\theta_3$ obtained in this way. For clarity, figure 4b focuses on the
region of $\sin^22\theta_3$ close to 1, which is the interesting one.
We can impose the experimental limits on these angles directly in this
figure. 
We see that the limit $\sin^22\theta_2<0.64$ (dashed line) translates into
a lower limit $\sin^22\theta_3\simgt 0.94$ (the region to the left of
the dashed line is forbidden). In a similar way, $\sin^22\theta_1>0.82$
(dotted lines) requires either $\sin^22\theta_3<0.09$ (but this solution
gives 
$\sin^22\theta_2>0.64$ and is not acceptable) or
$\sin^22\theta_3\simgt 0.9978$ (the region to the right of the dotted
line in figure 4b is allowed). Note that if we impose the stringent
condition $\sin^22\theta_3< 0.99$, derived from some fits to solar data as
discussed in the Introduction, then no viable solution would exist. 
Without such condition we find that this framework can accomodate the
angles required by the data with
\bea
\sin^22\theta_1\geq 0.82,\;\;\;
\sin^22\theta_2\leq 0.17,\;\;\;
\sin^22\theta_3\geq 0.9978.
\label{tetas2}
\eea
Since, in first approximation, the perturbed ${\cal M}_\nu$ will have
eigenvectors of the form (\ref{eigenv}), this result and the previous
discussion are also valid for the SM case \cite{cein} and any model
starting with ${\cal M}_\nu={\cal M}_b$. Actually, the results shown 
in eq.(\ref{tetas}) and Fig.~4 may be considered as
predictions of such a kind of scenarios. 

In the next section we show that there are regions in parameter space
where both the mass splittings and the mixing angles have the right values
to explain the solar and atmospheric neutrino data.

\subsection{Numerical Results}

\begin{figure}[t]
\centerline{\hbox{  
\psfig{figure=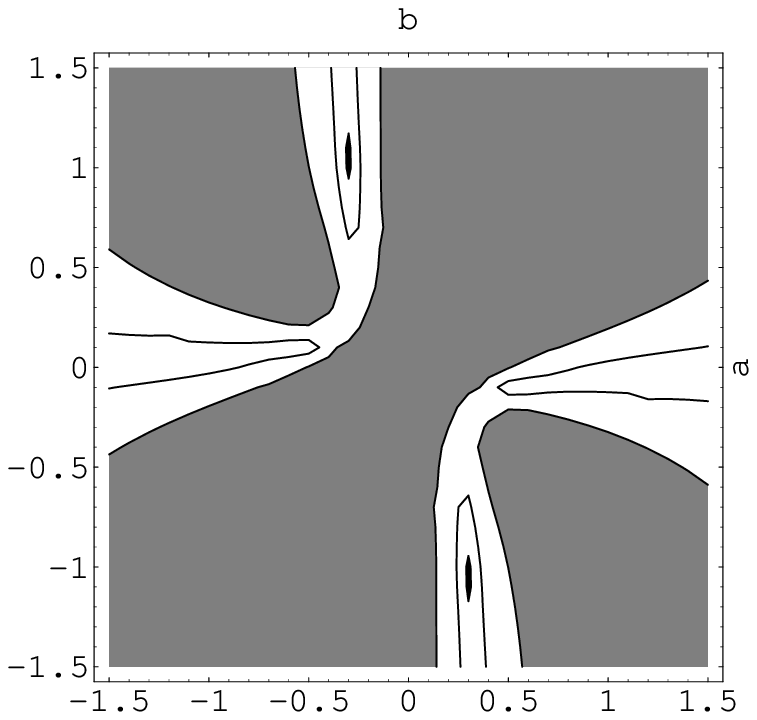,height=10cm,width=10cm,bbllx=1.cm,%
bblly=0.cm,bburx=12.cm,bbury=9.cm}
\psfig{figure=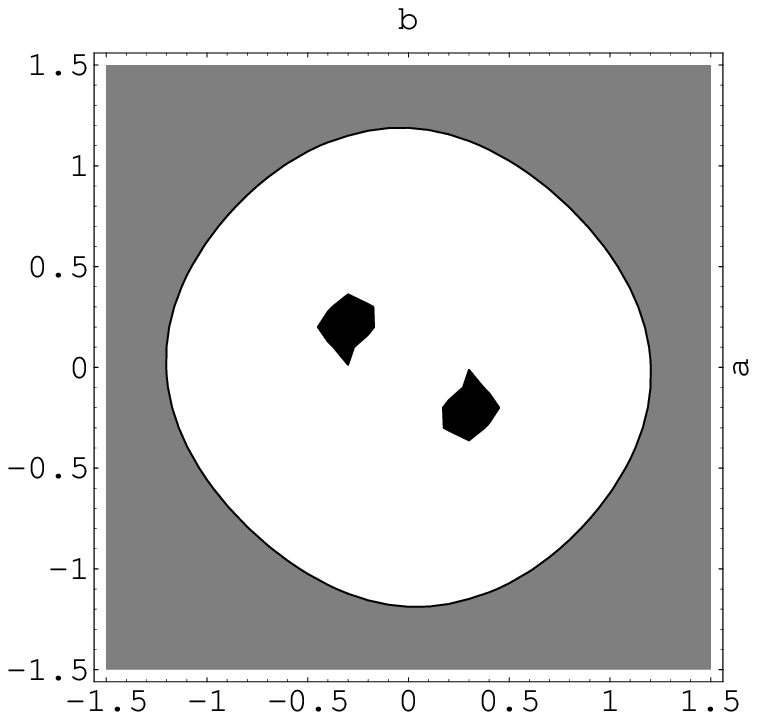,height=10cm,width=10cm,bbllx=4.cm,%
bblly=0.cm,bburx=15.cm,bbury=9.cm}}}
\caption{\footnotesize Left plot:
contours of $\Delta m_{12}^2/{\mathrm eV}^2$ in the $(b,a)$
plane from less than $10^{-5}$ (black area), through $6\times 10^{-5}$ 
(lines) to more than $2\times 10^{-4}$ (grey).
Right plot: same for $\Delta m_{23}^2/{\mathrm eV}^2$, from
$5\times 10^{-4}$ (black) to $10^{-2}$ (grey). The Majorana mass is
$10^{10}$ GeV and $\tan\beta=2$.}
\end{figure}

Figures 5 to 10 present our results for the mass splittings and mixing
angles at low energy, after numerical integration of the RGEs from $M_p$
to $M_Z$ as described in the previous sections. We stress that our 
convention in all the figures is to assign the indices 1 and 2 to the
neutrino states closest in mass, with
$m_{\nu_1}^2<m_{\nu_2}^2$ and leave the index 3 for the other neutrino
state. In the interesting regions, where the parameters
would fit well the solar and atmospheric data, the third neutrino 
is the heaviest. Hence, the neutrino spectrum has the same structure as in
the SM case~\cite{cein}.

We start with the results for $M=10^{10}$
GeV and $\tan\beta=2$ chosen as a typical example; 
the dependence of the results with $M$ and $\tan\beta$ is
discussed afterwards.
\begin{figure}[t]
\centerline{\hbox{
\psfig{figure=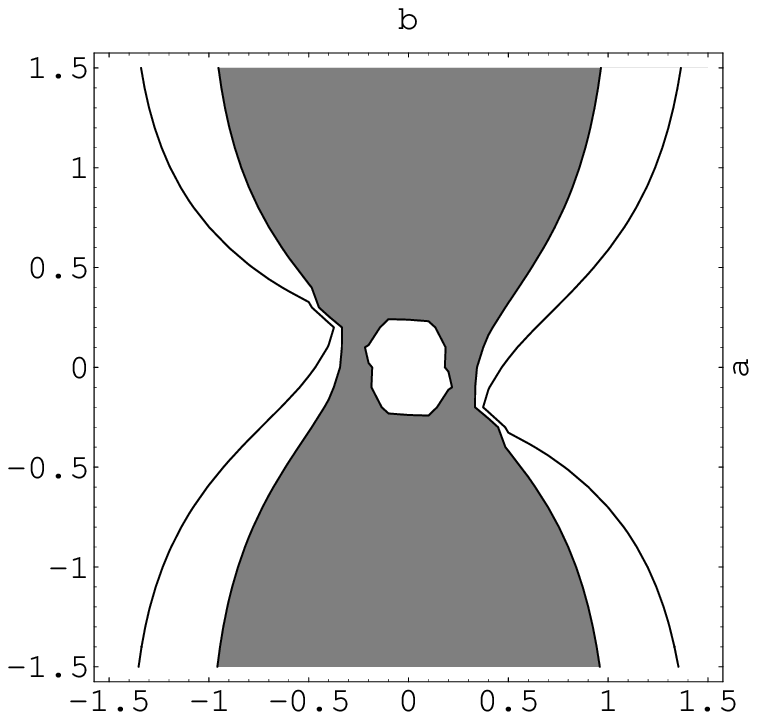,height=10cm,width=10cm,bbllx=1.cm,%
bblly=0.cm,bburx=12.cm,bbury=9.cm}
\psfig{figure=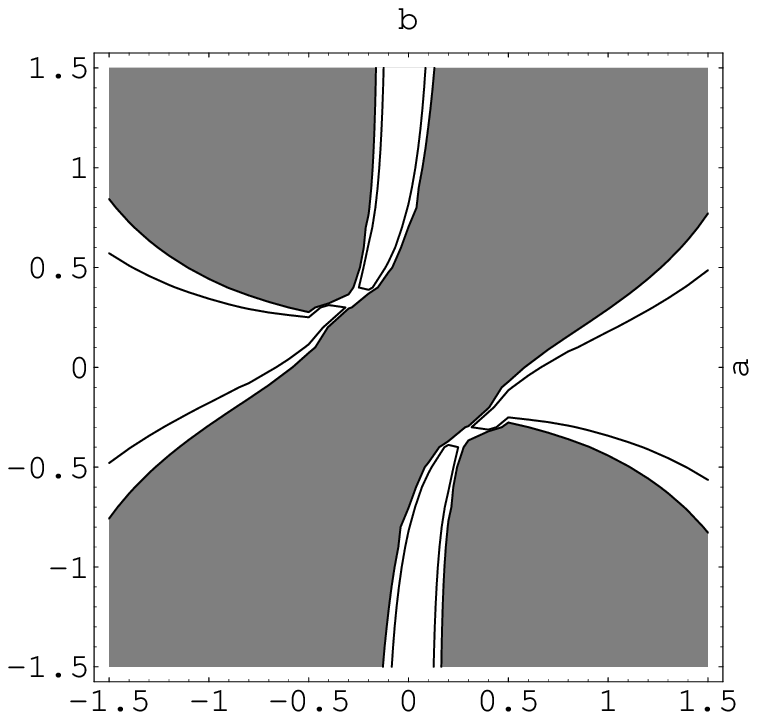,height=10cm,width=10cm,bbllx=4.cm,%
bblly=0.cm,bburx=15.cm,bbury=9.cm}}}
\caption{\footnotesize Left plot: Contours of
$\sin^22\theta_{2}$ in the $(b,a)$ plane. The grey area marks the
$\sin^22\theta_{2}>0.64$ region. The line singled-out corresponds to
$\sin^22\theta_{2}=0.36$. Right
plot: Contours of
$\sin^22\theta_{1}$ in the $(b,a)$ plane. In the grey area
$\sin^22\theta_{1}$ is smaller than 0.82, and the line corresponds to
$\sin^22\theta_{1}=0.9$. The Majorana mass is
$10^{10}$ GeV and $\tan\beta=2$.}
\end{figure}
Figure 5, left plot, shows contour lines of constant $\Delta m_{12}^2$
(the squared mass difference between the lightest neutrinos) in the plane
$(b,a)$. The black (grey) region is excluded because there $\Delta
m_{12}^2<10^{-5}\,
{\mathrm eV}^2$ ($\Delta m_{12}^2>2\times 10^{-4}\, {\mathrm eV}^2$),
which is too
small
(large)
to account for the oscillations of solar neutrinos (LAMSW solution). The
white area is thus the allowed region. The line in it corresponds to
$\Delta m_{12}^2=6\times 10^{-5}\, {\mathrm eV}^2$. Note how the allowed
area is not too large and extends around the parametric line
$1-c_{2a}-2\sqrt{2} s_{2a}s_b=0$, as explained in the previous section.

Figure 5, right plot, gives contour lines of constant $\Delta
m_{23}^2$. 
The black (grey) region is excluded
because there $\Delta m_{23}^2<5\times 10^{-4}\, {\mathrm eV}^2$ ($\Delta
m_{23}^2>
10^{-2}\, {\mathrm eV}^2$), which is too small (large) to account for the
oscillations
of atmospheric neutrinos. Again, the white area is allowed. The
small black areas
correspond in fact to the ``undecidable'' case discussed in the 
Introduction: they might be rescued  by unspecified extra effects.
We do not present a plot for $\Delta m_{13}^2$ because it can always be
inferred from $\Delta m_{23}^2$ and  $\Delta m_{12}^2$. Moreover, in the
interesting case, $\Delta m_{12}^2\ll \Delta m_{23}^2$, one has $\Delta
m_{13}^2 \simeq\Delta m_{23}^2$.

The intersection of the white areas in both plots is non-zero and would
give the allowed area concerning mass splittings. It is always the case
that the area surrounding the origin is excluded. There, the mass
differences are always of the same order, and follow the same pattern
discussed in section 2 ($\Delta m_{23}^2=2 \Delta m_{12}^2$). In any
case we conclude that, away from the origin, there is a non-zero
region of parameter space where $\Delta m_{23}^2\gg
\Delta m_{12}^2$, in accordance with the values required to explain the
solar and atmospheric neutrino anomalies simultaneously.

\begin{figure}[t]
\centerline{\hbox{
\psfig{figure=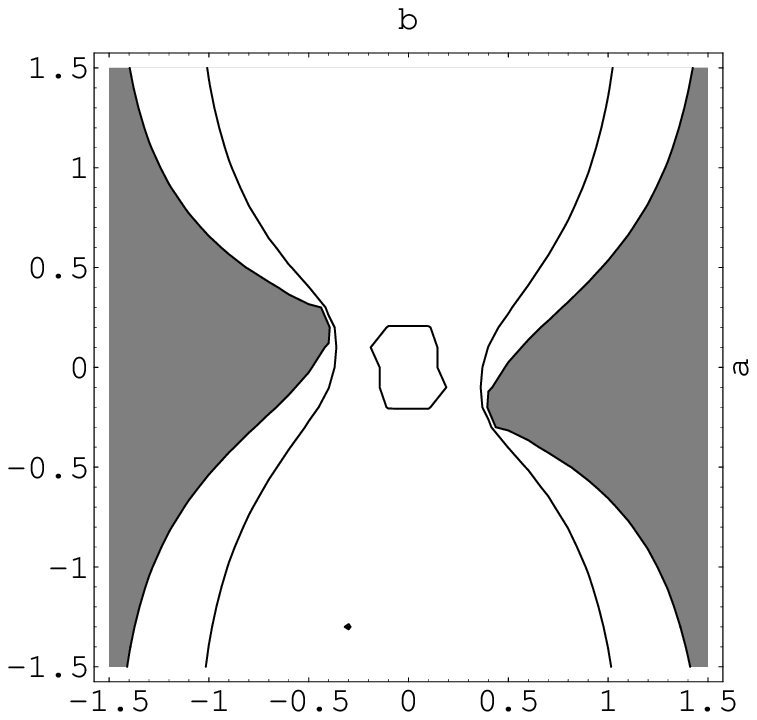,height=10cm,width=10cm,bbllx=1.cm,%
bblly=0.cm,bburx=12.cm,bbury=9.cm}
\psfig{figure=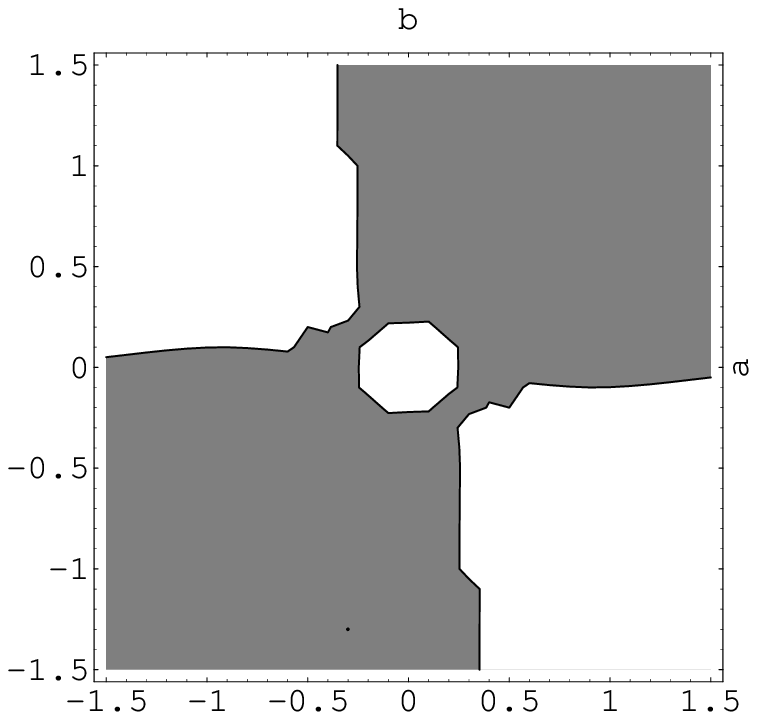,height=10cm,width=10cm,bbllx=4.cm,%
bblly=0.cm,bburx=15.cm,bbury=9.cm}}}
\caption{\footnotesize Left plot: Same as figure 6 for
$\sin^22\theta_{3}$. The grey area corresponds to values above 0.99.
The curves give $\sin^22\theta_{3}=0.95$. Right plot: The grey area 
corresponds to $\cos 2\theta_{3}<0$
}
\end{figure}

\begin{figure}[t]
\centerline{
\psfig{figure=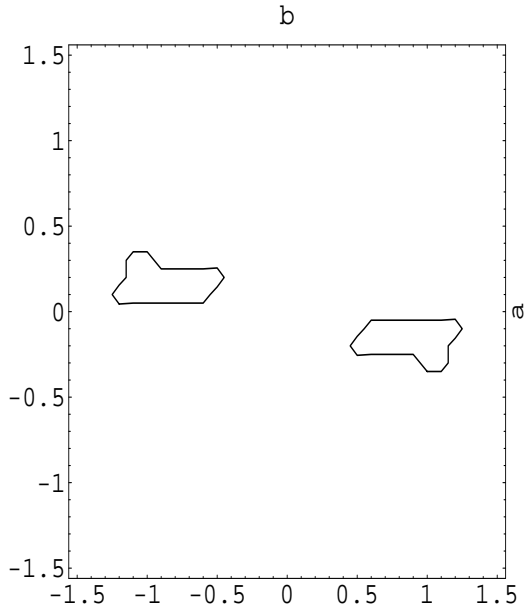,height=10cm,width=10cm,bbllx=2.cm,%
bblly=0.cm,bburx=13.cm,bbury=9.cm}}
\caption{\footnotesize Region (two disconnected parts)
in the $(b,a)$ parameter space for $M=10^{10}$ GeV and $\tan\beta=2$ where
all mass splittings and mixing angles 
satisfy experimental constraints. (See text for qualifications).
}
\end{figure}

\begin{figure}[t]
\centerline{\hbox{
\psfig{figure=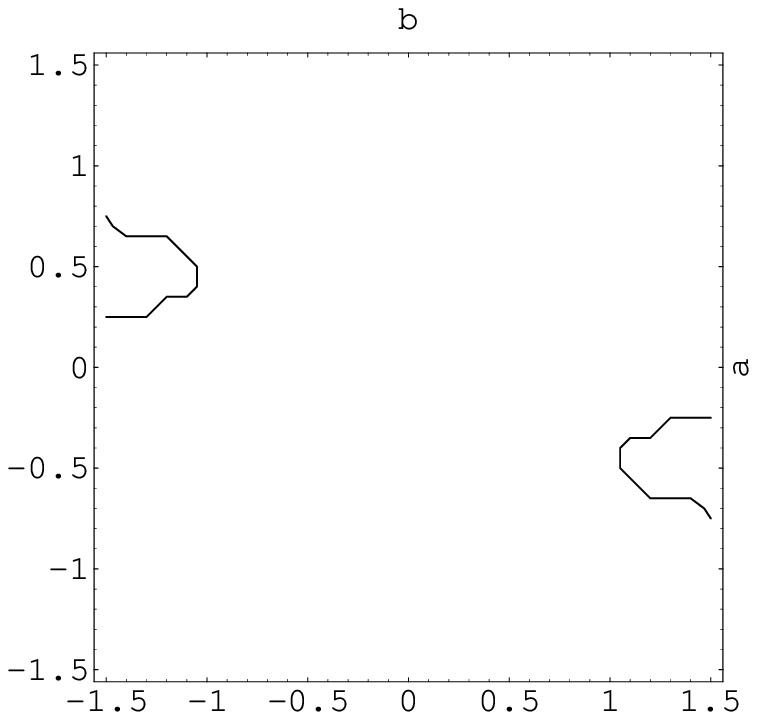,height=10cm,width=10cm,bbllx=1.cm,%
bblly=0.cm,bburx=12.cm,bbury=9.cm}
\psfig{figure=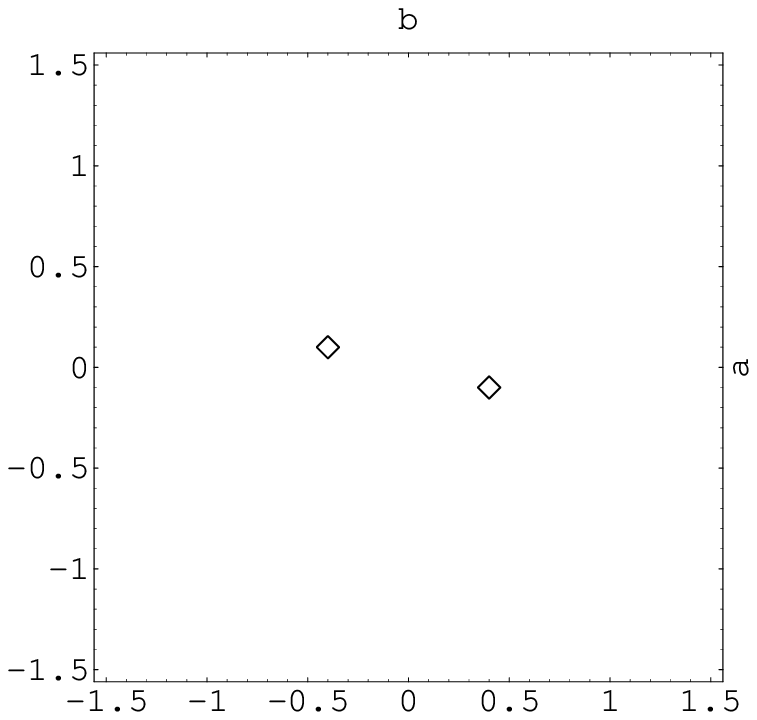,height=10cm,width=10cm,bbllx=4.cm,%
bblly=0.cm,bburx=15.cm,bbury=9.cm}}}
\caption
 \noindent{\footnotesize Same as figure 8 for
$\tan\beta=2$ and 
different values of the Majorana mass. Left plot: $M=10^{9}$ GeV; 
Right plot: $10^{11}$ GeV. 
}
\end{figure}

For mixing angles, figure 6, left plot, gives contours
of constant $\sin^22\theta_{2}$ (one of the mixing angles relevant for
atmospheric neutrino oscillations). The grey (white) area has
$\sin^22\theta_{2}$ larger (smaller) than 0.64 and is disfavored (favored)
by the data (SK $+$ CHOOZ) at 99\% C.L. according to the most recent 
analysis (last paper of ref.~\cite{range}).
The line singled-out corresponds to
$\sin^22\theta_{2}=0.36$ (maximum allowed value at 90\% C.L. according
to the same reference). Figure 6, right plot, shows contours of constant
$\sin^22\theta_{1}$ (the other mixing angle relevant for atmospheric
neutrinos). The grey (white) area
corresponds to $\sin^22\theta_{1}$ smaller (larger) than 0.82, and is
thus disallowed (allowed). The additional line included has
$\sin^22\theta_{1}=0.9$.

Finally, figure 7, left plot, presents contours of constant 
$\sin^22\theta_{3}$ which
is relevant for oscillations of solar neutrinos. The grey (white) region
has $\sin^22\theta_{3}$ larger (smaller) than 0.99. If one is willing to
interpret the existing data as implying an upper bound of 0.99 on
$\sin^22\theta_{3}$, then the grey region would be excluded. The plotted
curves give $\sin^22\theta_{3}=0.95$. 
Figure 7, right plot, shows the region of 
the parameter space accomplishing the resonance condition 
($\cos2\theta_{3}>0$), which is required for an efficient MSW solution
of the solar anomaly (see however the first paper of ref.~\cite{range} 
for caveats on this issue).

\begin{figure}
\centerline{\hbox{
\psfig{figure=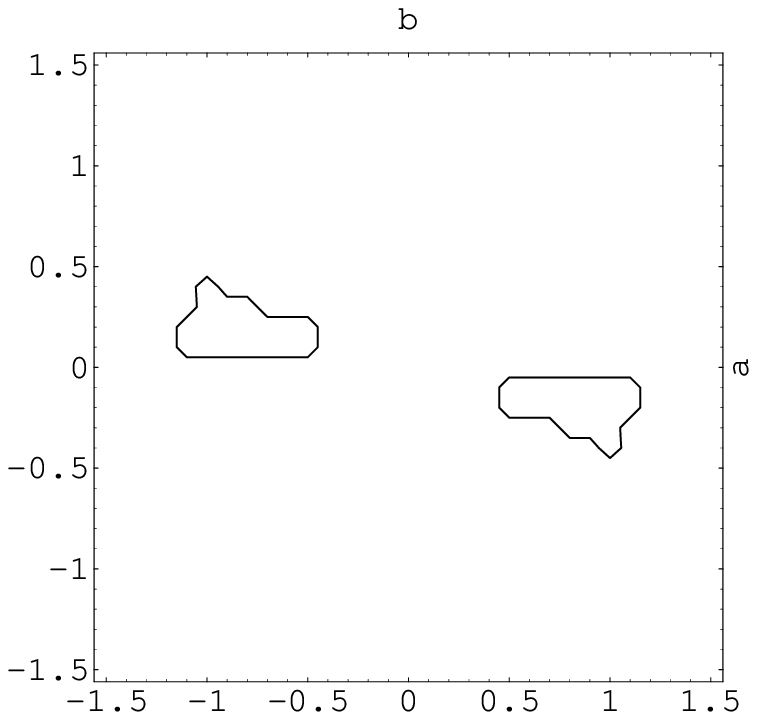,height=10cm,width=10cm,bbllx=1.cm,%
bblly=0.cm,bburx=12.cm,bbury=9.cm}
\psfig{figure=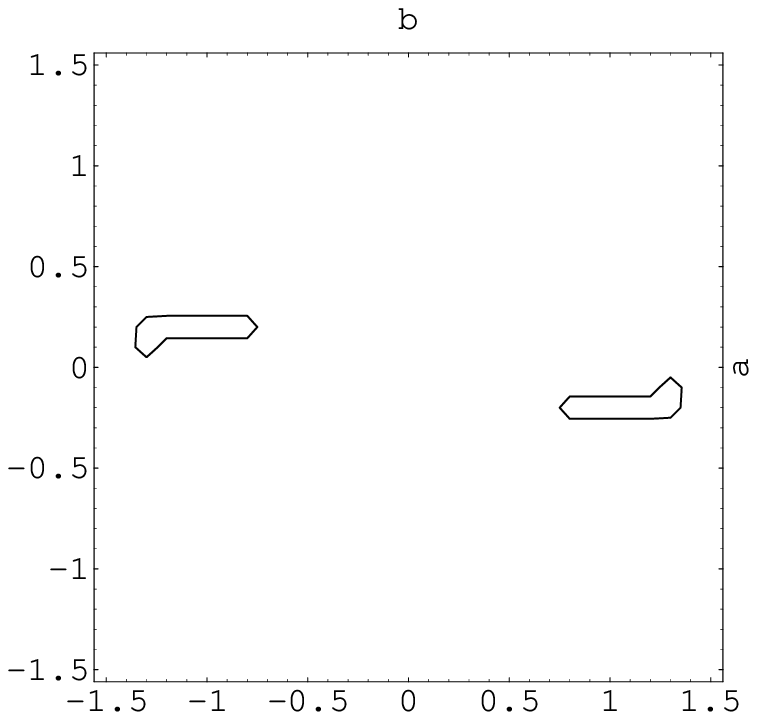,height=10cm,width=10cm,bbllx=4.cm,%
bblly=0.cm,bburx=15.cm,bbury=9.cm}}}
\centerline{\hbox{
\psfig{figure=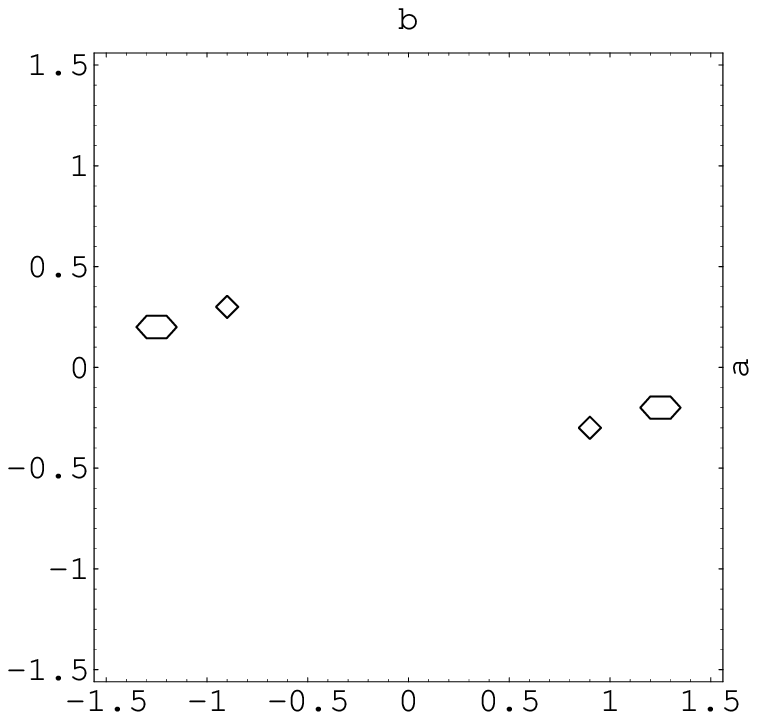,height=10cm,width=10cm,bbllx=1.cm,%
bblly=0.cm,bburx=12.cm,bbury=9.cm}
\psfig{figure=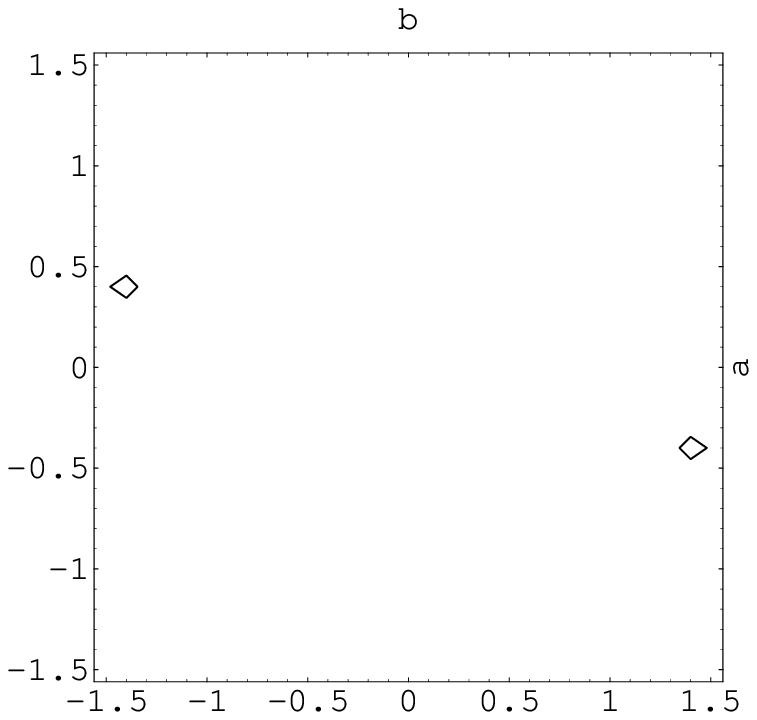,height=10cm,width=10cm,bbllx=4.cm,%
bblly=0.cm,bburx=15.cm,bbury=9.cm}}}
\caption
 \noindent{\footnotesize Same as figure 8 for
$M=10^{10}$ GeV  and 
different values of $\tan\beta$. Upper left: $\tan\beta=$ 1.75. Upper
right: 3.
Lower left: 4. Lower right: 6.5.}
\end{figure}

The region of parameter space where all constraints on mixing angles and
mass splittings are satisfied is given by the intersection of all white
areas in figures 5, 6 and 7 (right plot). 
If $\sin^22\theta_{3}<0.99$ is imposed, then
that intersection region, including now figure 7 (left plot), 
is empty and no allowed region remains. 
It should be noticed that this fact does not come from an incompatibility
between the previous constraint and the $\sin^22\theta_{3}>0.99$ obtained
from neutrinoless double $\beta$-decay limits, eq. (\ref{doublebeta}), in the
$\theta_2=0$ approximation. If this were
the case, it could be easily solved by decreasing the overall
size of the neutrino masses,  $m_\nu$, in eq.(\ref{doublebeta}), and this
is not the case. Indeed,  eq.(\ref{doublebeta}) is satisfied in nearly all
the parameter space. Even where $\sin^22\theta_{3}<0.99$, this is still true
thanks to  the contribution of $\theta_2$.  What actually
forbids the whole parameter space if $\sin^22\theta_{3}<0.99$ is imposed
is the incompatibility between acceptable $\theta_1$, $\theta_2$ and 
$\theta_3$ angles to fit simultaneously
all the neutrino oscillation data,
as can be seen from the figures and in agreement with the discussion
of the previous subsection, see eq.(\ref{tetas2}). 
This fact remains when $m_\nu$ is
decreased.  In fact, the effect of decreasing $m_\nu$ is essentially an
amplification  of the figures shown here, which comes from the fact that
for a given Majorana mass, the neutrino Yukawa couplings become smaller
(the effect is similar to decreasing $M$, which is discussed below).

If the $\sin^22\theta_{3}<0.99$
condition is relaxed (as discussed in the Introduction), 
then the allowed region is given by the two islands in figure
8, which is non-negligible. It is remarkable that the small regions where
all the mass splittings are of the right size and the small regions where
the mixing angles are the appropriate ones have a non-zero overlapping.

The dependence of the results with the Majorana mass $M$ is illustrated by
figure 9, where the allowed regions in the plane $(b,a)$ are shown for 
two different Majorana masses, $M=10^{9}$ and $10^{11}$ GeV, keeping
$\tan\beta=2$. For comparison with the allowed region for $M=10^{10}$ GeV
(given in figure 8) we see that for small $M$ the allowed region grows
and flies away from the origin, until it leaves the naturalness region
$|a|,|b|<1.5$. For  $M=10^8$ GeV no allowed region inside the natural range 
for $(a,b)$ remains. 
Conversely, increasing $M$ reduces the allowed region,
which gets closer to the origin (at $M=10^{12}$ GeV the allowed region
becomes extinct). 

Concerning the remaining parameter, $\tan\beta$, figure 10 gives 
the allowed regions for $M=10^{10}$ GeV and four different values of
$\tan\beta$: 1.75, 3, 4 and 6.5, as indicated. The minimum in $\tan\beta$
is dictated by the requirement of perturbativity of all couplings up to
the Planck scale, banning the presence of a Landau pole below it, as
discussed in section 5. The allowed region gets smaller and smaller when 
$\tan\beta$ increases. As explained in the previous section,
$|\epsilon_\tau|$ grows with $\tan\beta$ making harder and harder 
a cancellation that gives the correct $\Delta m^2_{sol}$ and
mixing angles. Eventually, for $\tan\beta\simgt 6.5$ the allowed region 
disappears (for $M=10^{11}$ GeV that value is $\tan\beta\simgt 3.5$). 

Finally, let us stress that, if the $\sin^22\theta_{3}<0.99$
condition is imposed, the whole parameter space becomes disallowed 
for any value of $M$ and $\tan\beta$. 
We also find that, whenever there is a hierarchy in the mass splittings,
the two lightest eigenvalues have opposite signs. This is just what is
needed to have a cancellation occurring in the neutrinoless double
$\beta$-decay constraint (\ref{doublebeta}). This constraint is satisfied in
almost the whole parameter space for any $M$.

\section{Examples of acceptable ans\"atze}

It is possible to find examples of matrices ${\bf Y_\nu}(M_p)$ which both 
fall inside the allowed areas presented in the previous section and 
are natural (perhaps pointing to a possible underlying symmetry).

An example, already presented in \cite{cein}, which still survives in the
supersymmetric case (for $M\sim 10^{10}\, {\rm GeV}$  and
$\tan\beta\sim 2$) is
\bea
{\bf Y_\nu}(M_p) =  Y_\nu\,\pmatrix{
- {\displaystyle{1\over 2\sqrt2}} &  1  &  1 \cr
 {\displaystyle{1\over 2\sqrt2}}  &  1  &  1 \cr
       0  &  
- {\displaystyle{1\over\sqrt2}} &  {\displaystyle{1\over\sqrt2}} \cr
}.
\label{ansatz}
\eea
It corresponds to $a=0$ and $b=\sinh^{-1}(3/4)\simeq 0.69$.
The mass splittings are
\bea
\Delta m_{12}^2\simeq 1\times 10^{-4}\, {\mathrm eV}^2,\;\;\;
\Delta m_{13}^2\simeq 2\times 10^{-3}\, {\mathrm eV}^2,\;\;\;
\Delta m_{23}^2\simeq 2\times 10^{-3}\, {\mathrm eV}^2,
\eea
and the mixing angles
\bea
\sin^22\theta_{2}=0.082,\;\;\;
\sin^22\theta_{1}=0.9163,\;\;\;
\sin^22\theta_{3}=0.99954,
\eea
with $\cos 2\theta_3$ at the border of the resonance condition for the
MSW mechanism.

Another examples of working ans\"atze can be obtained. 
For instance, the following ansatz (corresponding to
$a=-\cosh^{-1}(\sqrt{5}/2)
\simeq -0.48$, $b=\log(\sqrt{10}/2)\simeq 1.15$   )
\bea
{\bf Y_\nu}(M_p) =  Y_\nu\,\pmatrix{
- {\displaystyle {1\over 4}} &{\displaystyle{ 3\over \sqrt2}}   & 
{\displaystyle{ \sqrt2}}  \cr
 {\displaystyle{1\over 2\sqrt5}}  &  {\displaystyle{\sqrt5\over 2}}  &  
{\displaystyle{\sqrt5\over 2}} \cr
  {\displaystyle{1\over 4\sqrt5}}    &  - {\displaystyle{\sqrt5\over 2}} &
0 \cr
},
\label{ansatz2}
\eea
works correctly for 
$M\sim 10^{9}$ GeV and $\tan\beta\sim 2$, giving
\bea
\Delta m_{12}^2\simeq 1\times 10^{-4}\, {\mathrm eV}^2,\;\;\;
\Delta m_{13}^2\simeq 9\times 10^{-4}\, {\mathrm eV}^2,\;\;\;
\Delta m_{23}^2\simeq 8\times 10^{-4}\, {\mathrm eV}^2,
\eea
and the mixing angles
\bea
\sin^22\theta_{2}=0.02,\;\;\;
\sin^22\theta_{1}=0.979,\;\;\;
\sin^22\theta_{3}=0.99997,
\eea
with $\cos 2\theta_3>0$.
It could be interesting to explore possible symmetries that may be
responsible for the form of these ans\"atze and to analyze their
implications for future long-baseline experiments \cite{derujula}.

\section{Other relevant implications of 
neutrino-induced radiative corrections}

The previous sections focussed on the possibility of reproducing
the ``observed'' neutrino mass splittings from supersymmetric
radiative corrections. In particular we analyzed the case of nearly
degenerate neutrinos, which is closely related to the bimaximal mixing
scenario.

In this section, still working in a see-saw framework,
we take a different point of view and analyze the
physical impact of neutrino-induced radiative corrections. 
The results of this section are quite
generic.  They are not associated to the bimaximal mixing scenario,
not even to a scenario of degenerate neutrinos, but we will take this
case as a representative example to illustrate the phenomena. Some of
the following effects have already been mentioned along the paper, but
here they are studied in greater detail.

%

\vspace{0.2cm}
\noindent

The first topic concerns the appearance of Landau poles associated to the 
neutrinos and the corresponding implications.
It is easy to check from eq.(\ref{rg2}) that the neutrino Yukawa couplings,
${\bf Y_\nu}$, have a supersymmetric RGE quite similar to the top Yukawa
coupling. It is therefore not surprising that they can develop Landau poles
at high energy in a similar way. Obviously, the larger
 the right-handed neutrino Majorana mass, $M$, and the larger the low
energy neutrino masses, $m_\nu$, the larger the neutrino Yukawa couplings,
thus lowering the scale at which the Landau pole appears.  Consequently,
if, in order to preserve perturbation theory (and the nice supersymmetric
gauge coupling unification), we demand that the Landau pole does not appear
below $M_p$ (or the preferred high-energy scale), this puts upper bounds
on $M$ and $m_\nu$ (see ref.\cite{cdiq} for the analogue in the SM framework).

To analyze this effect, we will use the simplest textures of ${\bf
Y_\nu}$ and ${\cal M}$, leading to a bimaximal mass matrix, namely
${\bf Y_\nu}=Y_\nu {\bf I_3}$ and ${\cal M}(M_p)\propto  {\cal M}_{b}$
[${\cal M}$ has exactly the form given in eq.(\ref{MGG}) but with
overall scale $M$ instead of $m_\nu$]. Recall that this is equivalent
to the case $a=b=0$ in the analysis of textures performed in section
3. Obviously, for $a,b\neq 0$ the neutrino Yukawa couplings are 
larger and the corresponding bounds stronger. Hence, we are analyzing
here the most conservative case

\begin{figure}[t]  
\centerline{
\psfig{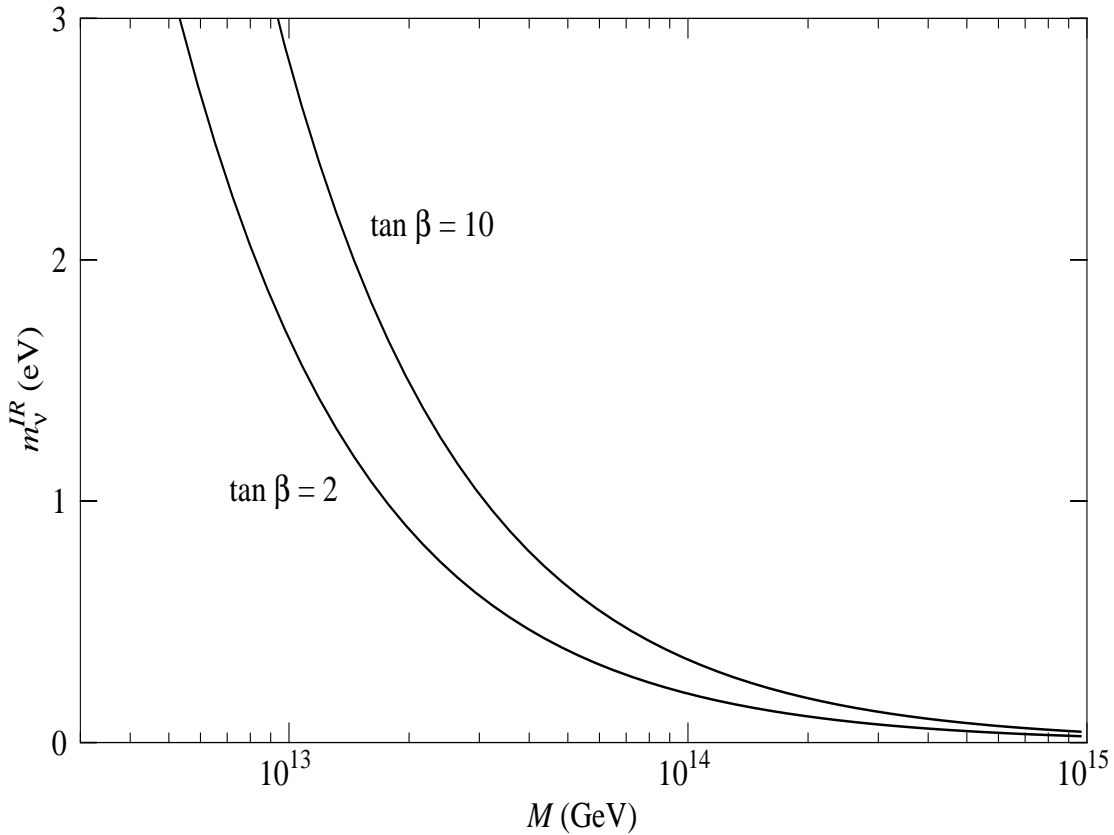}}
\caption
{\footnotesize 
Upper bound on the neutrino mass, $m_\nu^{IR}$, vs. the Majorana mass $M$ 
for two different values of $\tan\beta$.
}
\end{figure}
\begin{figure}[t]  
\centerline{
\psfig{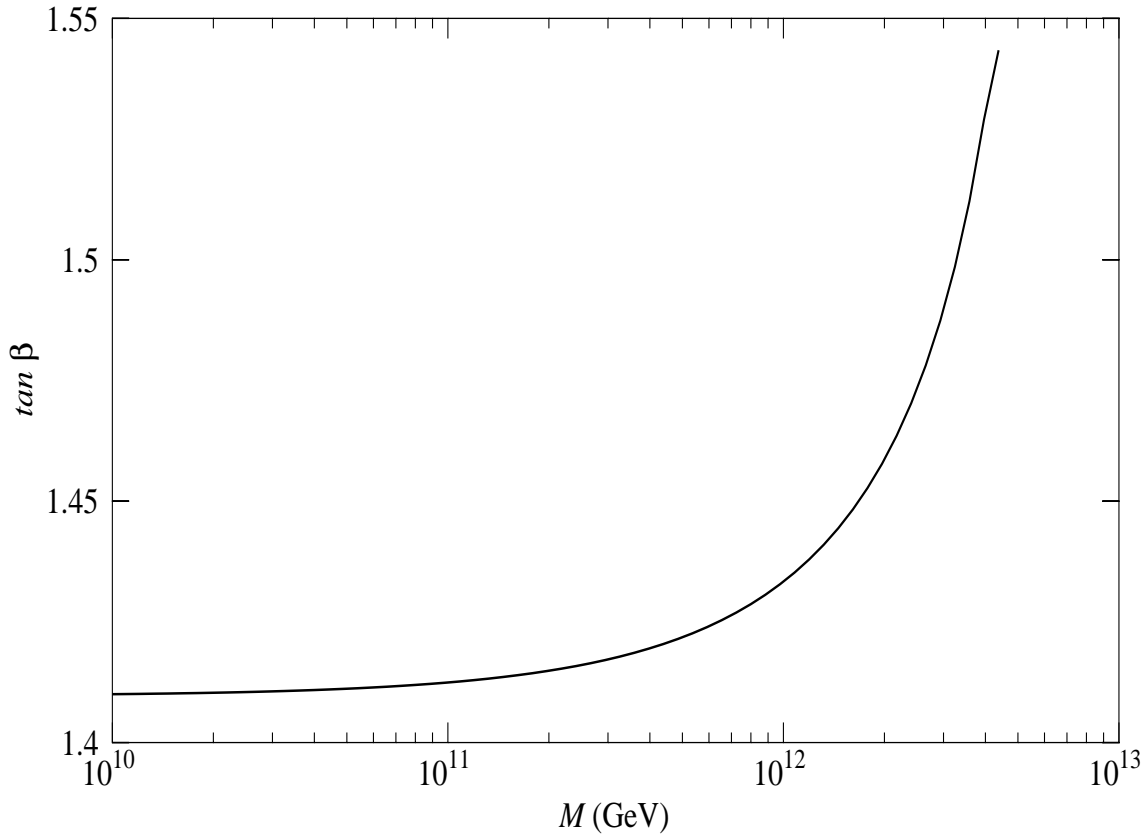}}
\caption
{\footnotesize 
Dependence of lowest value of $\tan\beta$ with the Majorana mass
for $m_\nu=2$ eV.
}
\end{figure}

To extract the bounds, we set the Landau pole of $Y_\nu$ at $M_{p}$
(i.e. $Y_\nu(M_{p})\gg 1$) and evaluate the corresponding low energy
value of $m_\nu$, through the renormalization group equations of
$Y_\nu$ (between $M_p$ and $M$) and $\kappa$ (below $M$), for a
certain value of the Majorana mass $M$.  This ``infrared fixed point''
value, say $m_\nu^{IR}$, represents an upper bound for the neutrino
mass.  The dependence of $m_\nu^{IR}$ on $M$ is illustrated in Fig.~11
for two different values of $\tan \beta$ (the value of $M$ in the
plot is to be understood as evaluated at the $M-$scale itself).
Alternatively, for a particular value of the neutrino mass, $m_\nu$,
we can extract from the figure the upper bound on the Majorana
mass. We note that the bounds are quite strong.  For quite moderate
values of the neutrino masses,  they conflict with the possibility of
a Majorana mass of ${\cal O}(M_{GUT})$.

%

\vspace{0.2cm}
\noindent
A different issue concerns the appearance of limits on $\tan\beta$.
In section 2, it was shown that the scenario of nearly degenerate neutrinos
is in conflict with large values of $\tan\beta$, the reason being that the
radiative corrections to the mass splittings become much larger than the
observed ones. The bounds were shown in Fig.3, for $m_\nu=2$ eV. For
instance, supposing that the scale at which the effective mass operator is
generated is $\Lambda=10^{11}$ GeV, it can be seen from that figure that an
upper limit $\tan\beta\simlt 7$ is obtained.  This was fully confirmed by
the see-saw analysis performed in section 3. As discussed in section 2,
these upper bounds become weaker as the neutrino mass, $m_\nu$, decreases.

Now, from completely different reasons, neutrino-induced radiative 
corrections modify the {\em lower} bound on $\tan \beta$.
In the ordinary MSSM the lowest
possible value for $\tan\beta$ is derived from the value of
the top Yukawa coupling in the infrared fixed point
(IFP) limit, by the condition of a physical top mass in agreement with the experiment.
In the MSSM extended with right-handed neutrinos, 
the  RGE for ${Y_t}$ becomes modified
in the form 
\bea 
\frac{d {\bf Y_U}}{dt}= -\frac{{\bf Y_U}}{16\pi^2}\left[\left(
\frac{13}{15}g_1^2+3 g_2^2+ \frac{16}{3} g_3^2- T_2
\right){\bf I}_3
- 3 {\bf Y_U^\dagger Y_U}- {\bf Y_D^\dagger Y_D} \right],
\label{YtRGE} 
\eea
(we write the RGE for the full matrix of up-quark Yukawa couplings).
If the neutrino Yukawa couplings, ${\bf
Y_\nu}$, are sizeable (which occurs for large enough $m_\nu$ or $M$), their
contribution tends to lower  the value of $Y_t$ further. In consequence,
$Y_t(m_{top})$ decreases and the corresponding $\tan\beta$ increases.
Fig.~12 shows the new lower bound on $\tan\beta$ for the typical case 
$m_\nu=2$ eV. Again, a diagonal structure for the neutrino Yukawa matrix,
${\bf Y_\nu}=Y_\nu {\bf I_3}$, has been assumed for simplicity.  It is
important
to realize that we cannot raise the value of $Y_\nu$ arbitrarily, since
then $Y_\nu$ develops a Landau pole below $M_p$, as has been discussed in
the previous subsection. This is the reason why the curve shown stops
abruptly.  As can be seen from the figure (evaluated at the 1-loop level), 
the change on $\tan\beta$ is
very modest, but it may give rise to an increment on the Higgs mass of
about 3 GeV, which should be taken into account for precision calculations.

\vspace{0.2cm}
\noindent
From a more qualitative point of view, we would like to mention two
important implications of the presence of massive neutrinos for the
supersymmetric perturbative unification. 

First, supersymmetric gauge coupling unification is considered as a
brilliant success of the MSSM, since the two-loop running of the
$SU(3)\times SU(2)\times U(1)$ gauge couplings unifies with great precision
at $M_{GUT}\sim 2\times 10^{16}$ GeV. The unification, although remarkable,
is not perfect. It is usual to invoke unknown (GUT or superstring)
threshold effects in order to explain the discrepance. Since neutrino
Yukawa couplings, ${\bf Y_\nu}$, give a 2-loop contribution to the $g_1$ and
$g_2$ gauge couplings (see e.g. \cite{pirogov}), for large enough ${\bf
Y_\nu}$ (which means large $m_\nu$ and/or $M$) this may be useful for a
complete satisfactory unification. 

Second, the ${\bf Y_\nu}$ couplings have a 1-loop contribution to the
charged {\em lepton} Yukawa couplings, ${\bf Y_e}$, see eq.(\ref{rg3}).
Again,
for large ${\bf Y_\nu}$, this produces significant variations for ${\bf
Y_e}$ at low energy, affecting the 
perturbative bottom-tau unification.
This subject has been already addressed in the literature \cite{botau}. 
Due to the form of eq.(\ref{rg3}), working in the flavour basis for the 
charged leptons, the mass eigenvalues $m_{e_i}$ receive a correction 
from ${\bf Y_\nu}$ that, at first order, goes like 
$\Delta_\nu m_{e_i}^2 \sim R\ m_{e_i}^2 
[{\bf Y_\nu^\dagger}{\bf Y_\nu}]_{ii}$, where $R$ is negative and
flavour-independent. Thus, at this order, the neutrino contribution
to the $m_b/m_\tau$ ratio goes always in the same sense for any
${\bf Y_\nu}$ texture. For the (problematic) first two families
this means in particular that neutrinos are not useful  
to rescue the analogous second generation ($m_s/m_\mu$) 
ratio, since the correction  goes in the 
wrong direction, but they might be useful for the $m_d/m_e$ ratio.
In any case, non-trivial ${\bf Y_\nu}$ textures will have a 
significant impact on the supersymmetric $m_b/m_\tau$ unification
scenarios.

\section{Conclusions}

We have studied the possibility that neutrinos masses have ${\cal O}(eV)$, 
in order to be cosmologically relevant. In that case they must be
nearly degenerate, as required by the oscillation interpretation 
of atmospheric and solar data. An important question is
whether radiative corrections have the right size to account
for the small mass splittings required or are generically too large for
the scenario to be considered natural. We addressed this problem in the
context of the SM (plus three right-handed neutrinos) in a previous
publication~\cite{cein} and, in this paper, we have extended this thorough
analysis to the supersymmetric case.

The size of the mass splittings that we find is always much larger than
required by the vacuum oscillation solution to the solar neutrino
problem, solution which is therefore excluded in this scenario.
  
When the origin of the non-zero neutrino masses is the see-saw mechanism
(we concentrate our study in this appealing case) we find non-negligible
regions in parameter space where the mass splittings are
consistent with the large angle MSW solution, providing a natural origin
for the $\Delta m^2_{sol} \ll \Delta m^2_{atm}$ hierarchy.
These regions correspond
to Majorana masses around $10^{10}$ GeV and small $\tan\beta\sim 2$.
Concerning the mixing angles, they are remarkably stable and close to the
bimaximal mixing form (something that is not guaranteed a priory,
due to an ambiguity in the diagonalization of the initial matrix).

We have understood analytically the origin of these remarkable features,     
giving explicit expressions for the mass splittings and the mixing angles.
In particular, we give simple analytical relations between the mixing
angles, which are also valid for any model starting with a bimaximal 
mixing form for the neutrinos.
In addition, we have presented particularly simple see-saw ans\"atze  
consistent with all atmospheric and solar neutrino observations.

Let us remark that the viability of the scenario is very sensitive to a   
possible upper bound on $\sin^2 2\theta_3$ (the angle responsible for the
solar neutrino oscillations). An upper bound
such as $\sin^2 2\theta_3<0.99$ would disallow completely the scenario
of nearly degenerate neutrinos due to the incompatibility between
acceptable mixing angles to fit simultaneously all the neutrino
oscillation data. This incompatibility is also easily understood
from the analytical relations between the mixing angles, and thus
is applies also for the SM case and more generic models.

Finally we have described in some detail several implications of the
existence of (possibly large) neutrino Yukawa couplings. This includes 
the effects on: the
triviality limits on the see-saw Majorana mass, the infrared fixed-point
of the top yukawa coupling, and the gauge and bottom-tau unification.

\section*{Acknowledgements}

This research was supported in part by the CICYT
(contract AEN95-0195) and the European Union
(contract CHRX-CT92-0004) (JAC). 
A.I. and  I.N. thank the CERN Theory Division for hospitality. 


\end{document}